\documentclass[useAMS,usenatbib,usegraphicx]{mn2e}
\usepackage{graphicx}
\usepackage[active]{srcltx}
\usepackage{times}
\usepackage[total={17.8cm,24.0cm},centering]{geometry}
\def\kms{$\mbox{km s}^{-1}$}
\newcommand{\sauron}{{\texttt {SAURON}}}
\newcommand{\Xsauron}{{\small {XSAURON}}}

\newcounter{subfigure}

\newcommand{\hb}{H$\beta$}
\newcommand{\hbp}{H$\beta^\prime$}
\newcommand{\mgb}{Mg\,$b$}
\newcommand{\mgbp}{Mg\,$b^\prime$}
\newcommand{\fes}{Fe5270$_{\rmn{S}}$}
\newcommand{\oiiia}{[O{\small III}]$\lambda 4959$}
\newcommand{\oiiib}{[O{\small III}]$\lambda 5007$}
\newcommand{\oiii}{[O{\small III}]}
\newcommand{\nifull}{[N{\small I}]$\lambda 5200$}
\newcommand{\nishort}{[N{\small I}]}

\newcommand{\re}{$(R_{\rmn e}/8)$}
\newcommand{\reb}{$R_{\rmn e}/8$}

\title[The SAURON project - VI.] {The SAURON project - VI.
  Line strength maps of 48 elliptical and lenticular galaxies}

\author[Kuntschner et al.] {Harald\ Kuntschner$^1$, Eric\ Emsellem$^2$,
  R.\ Bacon$^2$, M.\ Bureau$^{3}$, Michele Cappellari$^4$,
  \newauthor Roger\ L.\ Davies$^{3}$, P.\ T.\ de Zeeuw$^4$, Jes{\'u}s\ 
  Falc{\'o}n-Barroso$^{4}$, Davor\ Krajnovi{\'c}$^{3}$,
  \newauthor  Richard\ M.\ McDermid$^{4}$, Reynier\ F.\ Peletier$^{5}$ and Marc Sarzi$^3$ \\
$^1$Space Telescope European Coordinating Facility, European
  Southern Observatory, Karl-Schwarzschild-Str. 2, 85748
  Garching, Germany\\
$^2$CRAL-Observatoire, 9~Avenue Charles--Andr\'e,
    69230 Saint-Genis-Laval, France\\
$^3$Denys Wilkinson Building, University of Oxford, Keble Road, Oxford, 
     United Kingdom \\ 
$^4$Leiden Observatory, Postbus 9513, 2300~RA Leiden, The Netherlands\\
$^5$Kapteyn Astronomical Institute, Postbus 800, 9700 AV
     Groningen, The Netherlands}

\begin{document}
\maketitle
%
%
\begin{abstract}
  We present absorption line strength maps of $48$ representative
  elliptical and lenticular galaxies obtained as part of a survey of
  nearby galaxies using our custom-built integral-field spectrograph,
  \sauron, operating on the William Herschel Telescope. Using
  high-quality spectra, spatially binned to a constant signal-to-noise,
  we measure four key age, metallicity and abundance ratio sensitive
  indices from the Lick/IDS system over a two-dimensional field
  extending up to approximately one effective radius. A discussion of
  calibrations and offsets is given, as well as a description of error
  estimation and nebular emission correction. We modify the classical
  Fe5270 index to define a new index, \fes, which maximizes the usable
  spatial coverage of \sauron. Maps of \hb, Fe5015, \mgb\/ and \fes\/
  are presented for each galaxy. We use the maps to compute average
  line strengths integrated over circular apertures of one-eighth
  effective radius, and compare the resulting relations of index versus
  velocity dispersion with previous long-slit work. The metal
  line strength maps show generally negative gradients with increasing
  radius roughly consistent with the morphology of the light profiles.
  Remarkable deviations from this general trend exist, particularly the
  \mgb\/ isoindex contours appear to be flatter than the isophotes of
  the surface brightness for about 40\% of our galaxies without
  significant dust features. Generally these galaxies exhibit
  significant rotation. We infer from this that the fast-rotating
  component features a higher metallicity and/or an increased Mg/Fe
  ratio as compared to the galaxy as a whole. The \hb\/ maps are
  typically flat or show a mild positive outwards radial gradient,
  while a few galaxies show strong central peaks and/or elevated
  overall \hb-strength likely connected to recent star-formation
  activity. For the most prominent post-starburst galaxies even the
  metal line strength maps show a reversed gradient.
\end{abstract}

\section{INTRODUCTION}
\label{sec:intro}
We are carrying out a survey of the dynamics and stellar populations of
72 representative nearby early-type galaxies and spiral bulges based on
measurements of the two-dimensional kinematics and line strengths of
stars and gas with \sauron, a custom-built panoramic integral-field
spectrograph for the William Herschel Telescope, La Palma
\citep[hereafter Paper I]{bac01}. The goals and objectives of the
\sauron\ survey are described in \citet[][hereafter Paper II]{deZ02},
which also presents the definition of the sample. The full maps of the
stellar kinematics for the 48 elliptical (E) and lenticular (S0)
galaxies are given in \citet[][hereafter Paper~III]{em04}. The
morphology and kinematics of the ionised gas emission are presented in
\citet[][hereafter Paper V]{sar05}. The stellar and gaseous kinematics
of the spiral bulges are described in \citet{Fal05}. Here we present
maps of the absorption line strength measurements for the 48 E and S0
galaxies in the survey. The analysis of the line strength maps, and a
full analysis of the spiral galaxies in the sample will be presented in
later papers of this series. The data and maps presented here will be
made available via the \sauron\/ WEB page {\tt
  http://www.strw.leidenuniv.nl/sauron/}.

The measurement of absorption line strengths in combination with
stellar population models has been used for many years to probe the
luminosity-weighted age, metallicity and abundance ratios of certain
elements in integrated stellar populations
\citep[e.g.,][]{fab73,bur84,ros85,bro86,bic90,gon93,dav93,wor94,vaz99,TMB03}.
Integral-field spectroscopy allows to obtain spectroscopic information
over a contiguous area on the sky and thus to identify two-dimensional
structures \citep[see e.g.,][]{em96,pel99,dBur01}. For the first time
we can apply this technique to absorption line strength observations in
a large representative sample of early-type galaxies.

One of the most widely-used methods for investigating the optical
spectra of old ($\rmn{age} \ge 1$\,Gyr) integrated stellar populations
in early-type galaxies is to derive line strength indices in the
Lick/IDS system \citep{bur84,wor94b,tra98}, and compare them with
stellar population models. We adopt this system in order to allow
comparison with existing data. For future analysis we envisage the use
of higher resolution line strength systems once they become available.

In this paper we present the {\tt SAURON} line strength measurements
for our 48 representative E and S0 galaxies, together with a brief
analysis which focuses on the overall characteristics of the maps. A
full stellar population analysis of the line strength maps with the
help of stellar population models will be presented in a future paper
in this series.

This paper is organized as follows. In Section~\ref{sec:observations}
we summarize our observational campaign, while the data reduction steps
are outlined in Section~\ref{sec:reduction}, along with a discussion on
the flux-calibration of the data. In Section~\ref{sec:indices} we
describe the definition and measurement of the line strength indices.
The two-dimensional line strength maps are presented in
Section~\ref{sec:maps}, while in Section~\ref{sec:gra_ap} we discuss
line strength gradients and aperture corrections. In
Section~\ref{sec:central} we discuss average line strength measurements
extracted from circular apertures of one-eighth effective radius,
respectively, and compare them to the literature. The conclusions
follow in Section~\ref{sec:conclusions}. We comment on individual
galaxies in Appendix~\ref{sec:list}.

\section{Observations}
\label{sec:observations}

The \sauron\/ survey of 72 galaxies was carried out during eight
observing runs over four years (56 observing nights allocated). Details
on the observing conditions, instrument set-up and exposure times are
given in Paper~III. The \sauron\ sample of 48 elliptical (E) and
lenticular (S0) galaxies is representative of nearby bright early-type
galaxies ($cz\leq3000$\,\kms; $M_B\leq-18$\,mag). As discussed in
Paper~II, it contains $24$ galaxies in each of the E and S0 subclasses,
equally divided between `field' and `cluster' objects (the latter
defined as belonging to the Virgo cluster, the Coma~I cloud, and the
Leo~I group), uniformly covering the plane of ellipticity $\epsilon$
versus absolute blue magnitude $M_B$. Tables~A1 and A2 in Paper~II
provide basic information on the objects.

The \sauron\ survey is carried out with the low resolution mode of the
instrument, giving a field of view of $33\arcsec\times41\arcsec$,
contiguously sampled by 1431 $0\farcs94\times0\farcs94$ square lenses,
each of which produces a spectrum. Another 146 lenses sample a small
region $1\farcm9$ from the field center, which are used for
simultaneous sky subtraction. The spectral resolution of the observed
wavelength range 4800--5380\,\AA\ is $\sim$4.2\,\AA\ (FWHM,
$\sigma_{\rm inst} \simeq 108$\,\kms) and sampled at
1.1\,\AA\,pixel$^{-1}$. This wavelength range includes a set of
potential emission lines (e.g.\ \hb, [O{\small III}], [N{\small I}])
and stellar absorption lines (e.g.,\ \hb, Mg, Fe) which can be used to
investigate the stellar populations.

Each galaxy field was typically exposed for $4\times1800$\,s, dithered
by about 1\arcsec. In 30\% of the cases, we constructed mosaics of two
or three pointings to cover the galaxy out to about one effective
radius $R_{\rm e}$, or, for the largest objects, out to $\sim$$0.5
R_{\rm e}$.  The footprints of these pointings are shown overlaid on a
Digital Sky Survey image in Figure~1 of Paper~III.

In order to allow for inter-run calibration and provide templates for
redshift and velocity dispersion measurements we observed during each
run a number of stars covering a broad range of spectral types.
Specifically, we included stars from the Lick stellar library catalogue
\citep{wor94b} in order to calibrate our line strength measurements to
the Lick/IDS system and its associated models \citep[e.g.,][see also
Section~\ref{sec:offsets}]{wor94}. Spectrophotometric standard stars
were also observed to calibrate the response function of the system
(see Table~\ref{tab:flux_stds}).

\section{Data reduction}
\label{sec:reduction}
We reduced the \sauron\ observations with the dedicated \Xsauron\ 
software developed at CRAL--Observatoire, and described in Paper~I. The
basic reduction steps include bias and dark subtraction, extraction of
the spectra using a fitted mask model, wavelength calibration, low
frequency flat-fielding, cosmic-ray removal, homogenization of the
spectral resolution, sky subtraction, and flux calibration. The
wavelength calibration is accurate to $0.1$\,\AA\ ($6$\,\kms, rms).

Multiple exposures of the same galaxy were merged and mosaiced
together. In this process the wavelength domain was truncated to a
common range and the datacubes re-sampled to a spatial scale of
$0\farcs8 \times 0\farcs8$ with orientation North up and East to the
left. In order to allow for a meaningful analysis the final data-cubes
were spatially binned to a minimum S/N of 60 per \AA\/ using an
adaptive scheme developed by \citet[][]{cap03}. In this approach, the
spectra are co-added by starting from the highest S/N lenslet, and
accreting additional lenslets closest to the current bin centroid. A
new bin is started each time the target S/N is reached. The resulting
bin centroids are then used as starting points for a centroidal Voronoi
tessellation, ensuring compact non-overlapping bins and a uniform S/N
in faint regions. The binning used in this paper is identical to the
one used in Paper~III and V.

In the following we describe in more detail how the flux calibration
was derived, since this is an important issue for the derivation of
line strength indices which was not covered in detail in previous
papers of this series. We also summarize the removal of emission lines
which is another important step before the measurement of absorption
line strength.

\subsection{Flux calibration}
\subsubsection{Overall flux calibration}
During each observing run a number of spectrophotometric standards were
observed, typically close to the center of the field of view (hereafter
FoV) of \sauron\/ (see Table~\ref{tab:flux_stds}). As a first step we
evaluated the inter-run consistency of the spectral response function
of \sauron\/. For this purpose we investigated the relative changes in
the spectral response function in repeat observations of six flux
standard stars. In this analysis we are only interested in the relative
changes so we eliminated the overall throughput difference by
normalizing each set of observations of the same star in a 100\,\AA\/
region in the middle of the wavelength range. In total, 22 repeat
observations\footnote{Note, that some flux standard stars were observed
  multiple times within one run.} were used.

\begin{figure}
 \includegraphics[width=84mm]{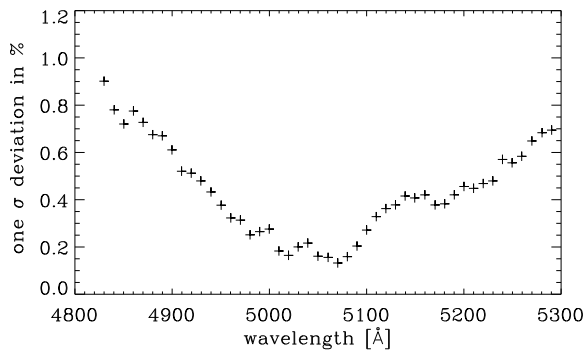}
 \caption{$1 \sigma$ standard deviation of the spectral response function 
   between our seven observing runs as a function of wavelength. The 22
   stellar observations used for this analysis were normalized in a
   100\,\AA\/ region at 5060\,\AA. The evaluation of the scatter was
   performed in $\sim$10\,\AA\/ steps.}
 \label{fig:flux_cal1}
\end{figure}

The results are presented in Figure~\ref{fig:flux_cal1}. The relative
run-to-run deviations in spectral response are smaller than 1\% ($1
\sigma$ standard deviation) over the full wavelength range. An analysis
of the variations in {\it absolute}\/ throughput for 13 standard stars
observed in photometric nights during the survey gives a standard
deviation of 5\%. No significant run-to-run variations could be
established. Since the overall stability of the instrument is
satisfactory we adopt a common flux calibration curve for all runs
which is described in the following.

In order to establish an absolute flux calibration for our spectra a
two step procedure was adopted: i) a relative spectral response
function was established which maps throughput variations on
$\sim$100\,\AA\/ scales introduced by a wavelength selection filter in
the instrument (see Paper~I); ii) the flux calibration tables compiled
by \citet[][``CALSPEC'' directory]{boh01} were used to establish an
absolute calibration of the \sauron\/ system.

The first step in the calibration procedure is necessary since
small-scale variations cannot easily be removed by the coarse
wavelength steps of classical flux calibration tables. Furthermore, it
is not uncommon to find large residuals at the position of absorption
features (e.g., \hb). To overcome these problems we observed in runs 4,
5, \& 7 the white dwarf EG\,131 (see Paper~III). This star shows an
exceptionally smooth continuum over the \sauron\/ wavelength range, and
in particular does not show any notable \hb\/ absorption (M.  Bessel,
private communication). This makes it an ideal calibration star to
remove small-scale continuum variations. Unfortunately, there is no
absolute flux calibration available for EG\,131, but the energy
distribution of a black-body with a temperature of $\sim$11800\,K can
be used to define the continuum (M.  Bessel, private communication).

\begin{figure}
 \includegraphics[width=84mm]{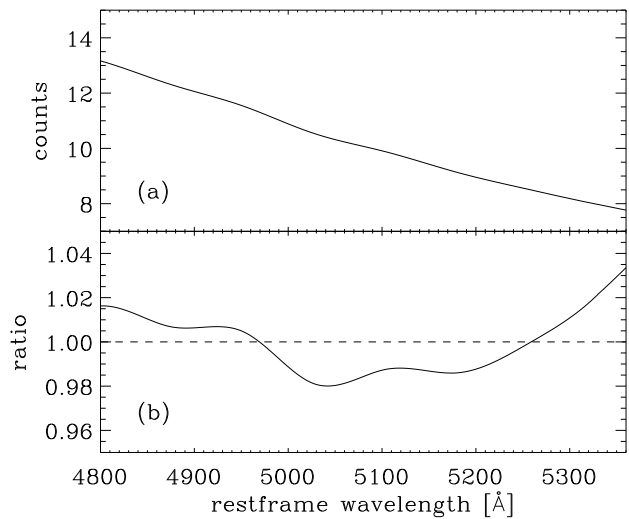}
 \caption{(a) The flux calibration curve for the \sauron\/ survey
   normalized to arbitrary units. (b) Small-scale variations of the
   flux calibration curve after removal of a linear fit.}
 \label{fig:flux_cal}
\end{figure}

Figure~\ref{fig:flux_cal} shows the result from our analysis of the
flux standard stars. We find several ``wiggles'' with a peak-to-peak
amplitude of up to 3\%. These wiggles, primarily derived from the
observations of EG\,131, can be confirmed in an independent way with
the standard star G191B2B since \citet{boh01} provide a model flux
calibration curve at a similar resolution as the \sauron\/
observations. This model flux curve is of very high quality and
provides accurate flux values even at small wavelength intervals.  Both
standard stars, EG\,131 and G191B2B give consistent results. In order
to establish the absolute flux calibration curve for the \sauron\/
survey we used all flux standard stars listed in
Table~\ref{tab:flux_stds} and the corresponding flux tables given in
\citet{boh01}.

\begin{table}
\centering
\begin{minipage}{75mm}
 \caption{List of spectrophotometric standard stars}
 \label{tab:flux_stds}
 \begin{tabular}{lll}
  \hline
Name     & Run observed & Calibration \\
 (1) & (2) & (3)\\
\hline               
Feige\,34 & 5,6,7     & STIS+Oke\\
Feige\,66 & 1,3       & FOS+Oke  \\
G191B2B   & 1,4,5,6   & Model\\
HZ\,44    & 1,3       & STIS \\
BD$+28$ 4211& 2,4     & STIS \\
BD$+33$ 2642& 7       & FOS+Oke  \\
HD\,09352 & 5         & STIS \\
EG\,131   & 4,5,7     & -    \\
%
\hline
 \end{tabular}

 \medskip 
 
 Notes: Listed are the spectrophotometric standard stars which were
 used to establish the flux calibration curve for the \sauron\/ system.
 Column~(1) lists the name of the star while column~(2) gives the run
 where the star was observed (see Paper~III). Column~(3) denotes the
 type of calibration file which was used for the absolute flux
 calibration \citep[see][for details]{boh01}. The star EG\,131 is a
 special continuum correction star. See text for details.

\end{minipage}
\end{table}

Before applying the overall flux calibration curve to our data the
observed spectrum is corrected for airmass according to the average
extinction curve for La Palma \citep{king85}. The spectra are not
corrected for Galactic extinction since the correction would be
negligible over the observed wavelength range.

Although we derive from the photometric standard star observations a
very stable behavior of the instrument and a well calibrated flux
curve, we cannot directly transfer this conclusion to the full data
set. For example, variations as a function of field of view and
sky-subtraction errors for low signal-to-noise regions can have a
significant effect on the flux calibration accuracy.

\subsubsection{Continuum correction for line strength indices}
\label{sec:cont}
\begin{figure}
 \includegraphics[width=84mm]{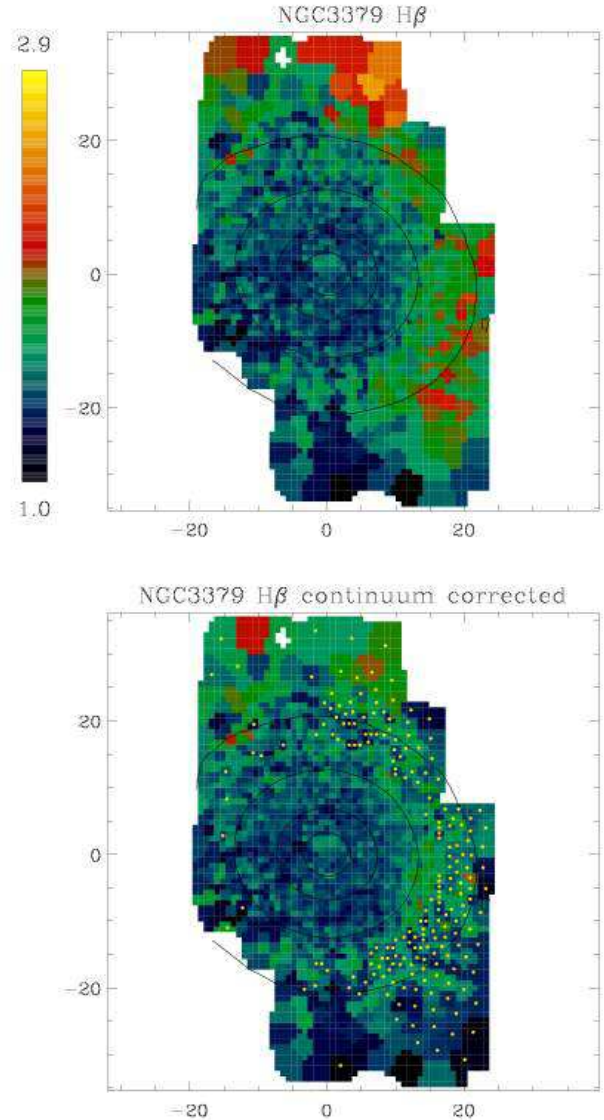}
 \caption{The \hb\/ map of NGC\,3379 before (top plot) and after
   (bottom plot) continuum correction. The range in \hb\/ strength is
   indicated by the color bar and the numbers to the top and bottom of
   the color bar give the range in \hb\/ [\AA]. The scale is the same
   for both plots.  The yellow dots in the bottom plot indicate all
   bins which were corrected for the continuum variation. North is up
   and east to the left. The spatial axis are given in arcsec.}
 \label{fig:ngc3379_cont}
\end{figure}

An analysis of our \hb\/ line strength measurements showed that we find
non-physically strong absorption strengths in the outer bins of some
galaxies often at the right side (as shown in
Figure~\ref{fig:ngc3379_cont}) of the FoV. A closer inspection revealed
that the abnormal \hb\/ absorption strength is mainly caused by a wrong
continuum shape across the \hb\/ region. In the outer bins typically
many individual lenslets need to be summed up in order to reach our
minimum target S/N of 60.  Therefore, any systematic errors, such as
imperfect tracing of the spectra on the CCD or sky subtraction, will be
present in the final binned spectrum.  Unfortunately we were unable to
remove these errors in our data reduction and thus decided to correct
the continuum in the binned spectrum.

In order to obtain a good template spectrum for each galaxy we derive
the optimal template for an average spectrum, within a 10\arcsec\/
radius, of each galaxy. This optimal template is the weighted
combination of 19 spectra, spanning a range in age and metallicity,
from the library of stellar population models of \citet{vaz99}, and
additionally six spectra from the \citet{jon97} stellar library from
which the Vazdekis models are built to provide spectra with large Mg
absorption strengths (for details see Paper~III). The optimal template
is obtained from an emission line corrected (see
Section~\ref{sec:emission_corr}) average spectrum, by simultaneously
fitting the stellar kinematics together with the templates, and using a
low-order polynomial.

For each individual bin of the galaxy in question we fit this optimal
template, with the kinematics fixed to the values of Paper~III,
together with an 11$^{th}$ order multiplicative polynomial continuum
over the full wavelength range of the spectrum. Then in a second step
we check if the continuum fit shows significant second and higher order
structure over the wavelength range of the \hb\/ index.  Linear
continuum variation does not affect the measured line strengths.  Only
for bins covering more than one lenslet and where we detect a
significant\footnote{The rms deviations are greater than 0.006 after
  the removal of a straight line fit.} continuum variation we divide by
the continuum fit and thus apply the continuum correction before
measuring the \hb\/ index.  All bins which have been continuum
corrected show a flag in the data-cubes of the public release and
furthermore, we added (in quadrature) a constant systematic error of
0.1\,\AA\/ to the \hb\/ index of corrected bins.  The typical fraction
of corrected bins in a data-cube is 4\% while the extremes go from 1\%
to 33\%. As an example Figure~\ref{fig:ngc3379_cont} shows the full
\hb\/ map of NGC\,3379 before and after the continuum correction.

\begin{figure}
  \includegraphics[width=84mm]{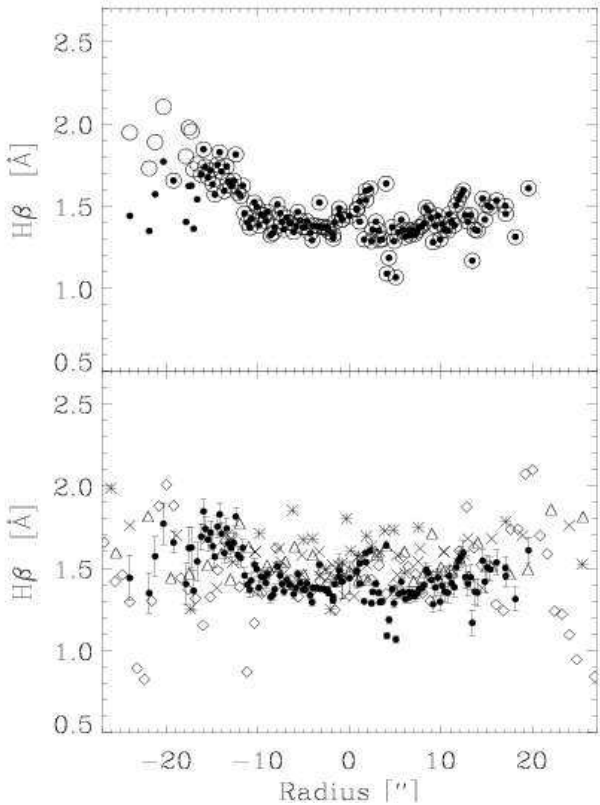}
 \caption{Effects of continuum correction for NGC3379 (major axis). The 
   top plot shows {\tt SAURON} data before (open circles) and after
   (filled circles) continuum correction for the \hb\/ index. The
   bottom panel shows a comparison with literature data. Error bars are
   only plotted for the {\tt SAURON}\/ data. The data of
   \citet[][]{dav93} is shown as crosses and mirrored about the zero
   point; data from \citet[][and private communication]{gon93} is shown
   as open triangles; the data of \citet[][]{rmd05} is shown as stars.
   \citet{san04} observations are shown as open diamonds. Literature
   data outside $\pm27\arcsec$ is not shown.}
 \label{fig:n3379_comp}
\end{figure}

In Figure~\ref{fig:n3379_comp} we show a simulated long-slit
observation of the \hb\/ absorption strength along the major axis of
NGC\,3379 derived from the \sauron\/ data before and after the
continuum correction. The top panel clearly shows the points at large
negative radii where the continuum has been corrected. The literature
comparison (bottom panel) shows a good agreement.

The success of our continuum correction is critically dependent on the
ability of the optimal template to fit the observed spectrum. In
general this may be a problem if the galaxy shows non-solar abundance
ratios which are not mimicked by the template library. However, in the
case of \hb\/ the index is only weakly dependent on abundance ratio
\citep{tra00a,TMB03} and thus the method works well. Continuum
variations can in principle also affect other indices such as Fe5015
and \mgb. Maps of Fe5015 are clearly affected, while \mgb\/ maps show
evidence of continuum problems for only a handful of galaxies. For
metal indices abundance ratios are a critical issue. The continuum
variations and unaccounted abundance ratio differences between the
galaxy spectrum and the optimal template are largely degenerate (at
least for our data and models). Thus we decided not to correct the
Fe5015, \mgb\/ and \fes\/ indices. For selected galaxies we remove bins
which suffer from significant continuum problems (see also
Section~\ref{sec:maps}).

\subsection[]{Emission-line corrections}
\label{sec:emission_corr}
Many early-type galaxies show signs of relatively weak emission from
ionized gas and dust in the interstellar medium
\citep[e.g.,][]{goud94}. Three out of four of the main absorption line
features in the \sauron\/ wavelength range are potentially affected by
emission lines; namely, \hb, \oiiia, \oiiib, and \nifull. Before
measuring {\it absorption line}\/ indices these emission lines, where
present, need to be removed taking into account that the gas kinematics
can be decoupled from the stellar kinematics. This is done as part of
our efforts to analyze the emission lines themselves. The emission-line
measurements and results were described in detail in Paper~V. Here we
give only a short summary.

For five galaxies we find no emission above our detection limit
(NGC\,821, NGC\,2695, NGC\,4387, NGC\,4564, and NGC\,5308); for a
further seven galaxies we find only weak evidence for emission
(NGC\,4270, NGC\,4382, NGC\,4458, NGC\,4473, NGC\,4621, NGC\,4660, and
NGC\,5845), while all of the remaining galaxies in our survey show
clear signs of emission. All emission lines are modeled with a simple
Gaussian profile. Using the stellar kinematics derived in Paper~III we
re-derive the best fitting, optimal stellar template\footnote{The
  composition of the template library was improved with respect to the
  one used in Paper~III. Three stars from the \citet{jon97} library
  were exchanged with more suitable ones which allowed a better match
  of the spectral characteristics in large early-type galaxies and thus
  an improved emission correction. The effects on the derivation of the
  stellar kinematics are negligible.} in combination with the best
fitting emission template featuring \oiii\/ only. While the \oiii\/
strengths and kinematics are free parameters in this first step, the
regions of potential \hb\/ and \nishort\/ emission are masked. If no
significant \oiii\/ emission is found we assume that there is also no
\hb\/ and \nishort. This limits our ability to measure weak emission
associated with star-formation regions where \hb\ / \oiii ~$\gg 1$.
This is an unavoidable consequence of the fact that {\em weak}\/ \hb\/
emission cannot always reliably be measured without constraining its
kinematics.  Whether emission is found or not is determined by a
measure of the amplitude of the emission lines with respect to the
level of both statistical and systematic deviations from the fit to the
stellar continuum. As a cutoff we use an amplitude-over-noise
(hereafter A/N) ratio of 4. This corresponds to a detection limit of
0.1\,\AA\/ in the central, high S/N, parts of the line strengths maps
and is reduced to approximately 0.2\,\AA\/ in the outer parts.

If we find significant \oiii\/ emission we re-fit the galaxy spectrum
allowing for potential \hb\/ and \nishort\/ emission. For galaxies with
weak emission lines we constrain the \hb\/ and \nishort\/ emission
kinematics to be the same as the \oiii\/ ones. For ten galaxies
(NGC\,2768, NGC\,3032, NGC\,3414, NGC\,4278, NGC\,4374, NGC\,4459,
NGC\,4486, NGC\,4526, NGC\,5838, and NGC\,5846) with relatively strong
emission we are able to fit the \hb\/ emission kinematics independently
while the \nishort\/ kinematics are tied to \hb\/ (see Paper~V for
details). The A/N detection limits for \hb\/ is set to 3 and 5 for the
constrained and unconstrained kinematics, respectively.  The detection
limits for \hb\/ emission are 0.06\,\AA\/ and 0.2\,\AA\/ in the inner
and outer parts of the galaxies, respectively.

For \nishort\/ the A/N detection limit is always set to 4. We note that
especially for the \nishort\/ emission line the dominating source of
noise is not shot noise from the spectrum, but template mismatch due to
unaccounted for non-solar abundance ratios in the galaxies. We
therefore measure the A/N ratio for the \nishort\/ emission line in the
\mgb\/ region which is a good guide for the degree of template
mismatch. Due to the difficulty of quantifying the effects of template
mismatch, the detection limits of the \nishort\/ line are not well
established. They will be, however, similar to the \oiii\/ detection
limits and somewhat larger for objects with large non-solar abundance
ratios.

\begin{figure}
 \includegraphics[width=84mm]{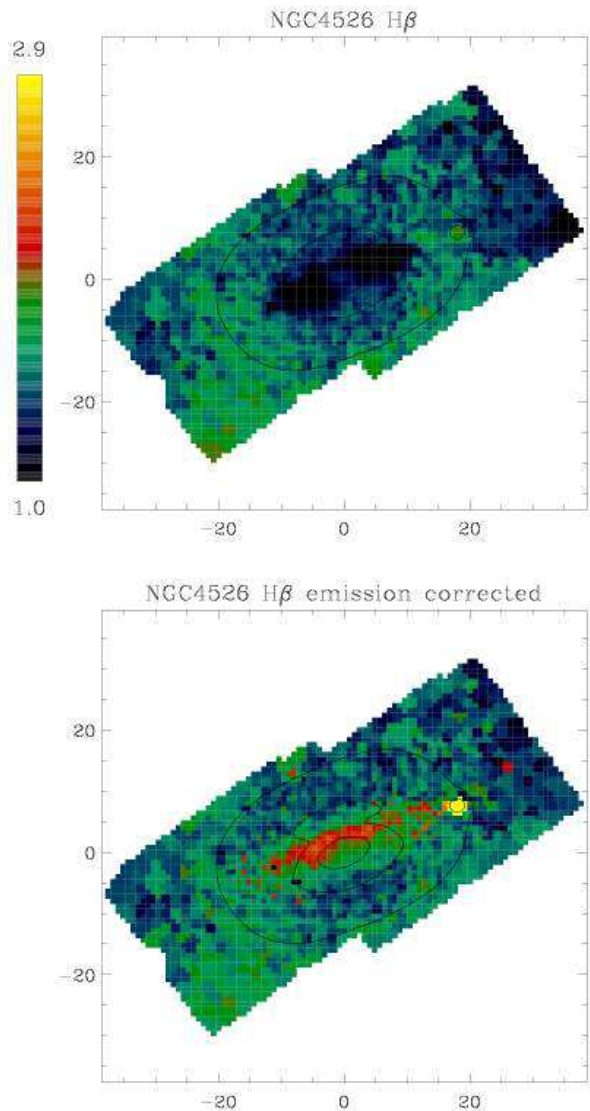}
 \caption{The \hb\/ map of NGC\,4526 before (top plot) and after
   (bottom plot) emission and continuum correction. The range in \hb\/
   strength is indicated by the color bar and the numbers to the top
   and bottom of the color bar give the range in \hb\/ [\AA]. The scale
   is the same for both plots. North is up and east to the left. The
   spatial axis are given in arcsec. The small circular region of
   strong \hb\/ absorption visible in both maps (Ra $\simeq$
   18\arcsec\/ and Dec $\simeq$ 7\arcsec) corresponds to the position
   of a star in the field of view.}
 \label{fig:em_corr}
\end{figure}

In Figure~\ref{fig:em_corr} we show an example map for the emission
correction of the \hb\/ feature. NGC\,4526 has an inclined dust disk in
the central $20\arcsec \times 20 \arcsec$ (see Paper~V) showing very
regular gas kinematics and a circumnuclear region with particularly
strong \hb\/ emission. After emission correction a relatively thin
elongated region of strong \hb\/ absorption coinciding with the dust
disk appears. This example is representative of the strongest emission
corrections in our data.

\begin{figure}
 \includegraphics[width=84mm]{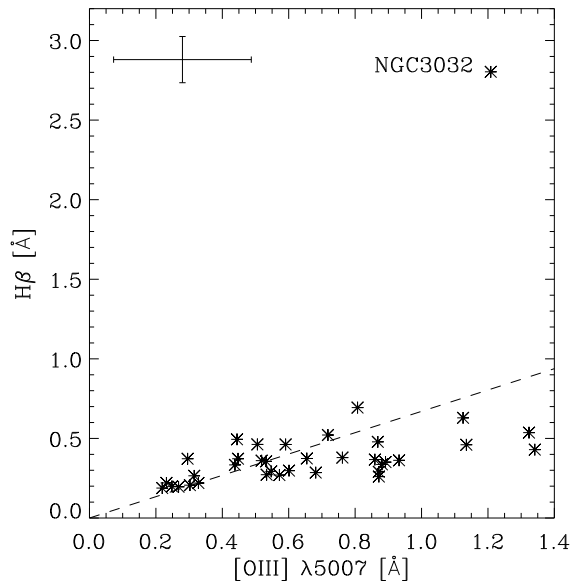}
 \caption{Median \hb\/ emission {\em vs}\/ \oiii\/ emission for all
   galaxies with significant detection of emission lines. The error bar
   in the top left corner shows a typical range of emission strength
   found {\em within}\/ individual galaxies (estimated as a robust $1
   \sigma$ standard deviation). The dashed line reflects the relation
   of \hb\/ = 0.67~$\times$~\oiiib\/ corresponding to our median
   correction factor. The position of NGC\,3032 is indicated by its
   name; for details see text. \looseness=-2}
 \label{fig:o3hb}
\end{figure}

Many previous authors corrected for \hb\/ emission using an average
ratio between the \oiiib\/ and \hb\/ lines. For instance \citet{tra00a}
concluded that 0.6 times the \oiiib\/ emission is a reasonable estimate
for the \hb\/ emission. However, in their sample of 27 galaxies the
correction factor varies from 0.33 to 1.25 and it is therefore doubtful
whether this correction is accurate for an individual galaxy \citep[see
also][]{meh2000}. We confirm this with our data (see
Figure~\ref{fig:o3hb}) and find a median correction factor of 0.67 with
a range from 0.3 to 2.3. There is weak evidence that the average
correction factor is a function of emission strength, becoming smaller
at large \oiiib\/ strength. More importantly, we point out that
emission line ratios generally vary significantly across the galaxies
(see error bar in Figure~\ref{fig:o3hb} and \oiii /\hb\/ maps in
Paper~V). NGC\,3032 shows an exceptionally large ratio of \hb /
\oiii\/ $\simeq 2.1$, pointing towards ongoing star-formation over a
significant area of the galaxy.

\section[]{The line strength measurements}
\label{sec:indices}
Here we describe which line strengths indices we use for the \sauron\/
survey, the calibration of our measurements to the standard Lick/IDS
system and the evaluation of the line strength errors.

\subsection[]{Line strength indices}
\begin{figure}
 \includegraphics[width=84mm]{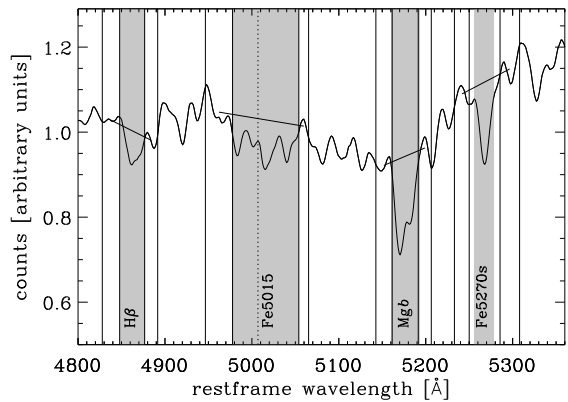}
 \caption{The set of line strength indices shown on a central
   \sauron\/ spectrum of NGC\,3379. The shaded regions are the central
   band-passes whereas the dashed lines indicate the side-band-passes.
   The solid lines show the pseudo-continuum determined from the
   side-band-passes. The location of the \oiiib\/ emission line is
   indicated by the dotted line.}
 \label{fig:lick_bands}
\end{figure}

In the Lick/IDS system absorption line strengths are measured by
indices, where a central feature bandpass is flanked to the blue and
red by pseudo-continuum band-passes (see Figure~\ref{fig:lick_bands}).
The mean height in each of the two pseudo-continuum regions is
determined on either side of the feature bandpass, and a straight line
is drawn through the midpoint of each one. The difference in flux
between this line and the observed spectrum within the feature bandpass
determines the index \citep[][]{tra98}. For most absorption features,
the indices are expressed in angstroms of equivalent
widths\footnote{For broad molecular bands, the index is expressed in
  magnitudes. No such indices are used in the \sauron\/ system.}.

The exact wavelength definitions of the indices, the observed
wavelength range and the redshift of the object determine which indices
can be measured. The full wavelength range of \sauron\/ is
4760--5400\,\AA. However, the tilting of the wavelength selection
filter in order to avoid ghost images, reduces the usable wavelength
range common to all lenslets to approximately 4825--5275\,\AA\/ (see
Paper~I). For the nearby sample of \sauron\/ galaxies (i.e., $cz \le
3000$\,\kms), this wavelength range allows us to measure three Lick/IDS
indices: \hb, Fe5015 and \mgb\/ (see Table~\ref{tab:lst_def}).

The usable wavelength range is a well-determined function of position
on the CCD. The red wavelength limit is most affected, and varies from
5275 to 5380~\AA\/ over the FoV.
For lenslets with a red-extended wavelength coverage we can measure a
further Lick/IDS index: Fe5270. It is desirable to measure this index
since it is an important ingredient to investigate the Mg-to-Fe
abundance ratios in integrated stellar populations \citep{wor92}.

Since the red wavelength cutoff is a monotonic function across the
field of view, by making the red bandpass of the Fe5270 index as blue
as possible, we effectively increase the usable field of view. For this
reason we have designed a new index for the Fe5270 absorption feature
with a reduced red wavelength coverage. The bandpass definitions of
all indices used for the \sauron\/ survey are summarized in
Table~\ref{tab:lst_def}.

The main aim of the re-definition of the Fe5270 index is to increase
the effective field of view. A useful re-definition of the index should
also, however, obey the following conditions: (i) small and
well-determined line of-sight-velocity-distribution (hereafter LOSVD)
corrections (see Section~\ref{sec:veldispcorr}); (ii) good Poisson
statistics; (iii) similar sensitivity towards changes in age,
metallicity and abundance ratios as the original Lick/IDS Fe5270 index
in order to allow easy conversion.

In order to evaluate the optimal new index definition we first created
a catalogue of model spectra resembling typical \sauron\/ observations
of early-type galaxies. In fact we employed the same model spectra
library as used for the optimal template determination (see Paper~III)
and the LOSVD corrections (see Section~\ref{sec:veldispcorr}). By using
this model library we cover reasonably well the range of expected
observations for the Fe5270 absorption feature.

All six wavelength values (see Table~\ref{tab:lst_def}) which define an
index were systematically changed in 0.5\,\AA\/ steps until an optimal
definition\footnote{The index definition was optimized at the
  instrumental resolution of \sauron\/ ($\approx 108$\,\kms) in order
  to allow for the new generation of stellar population models which
  will be able to make predictions at this spectral resolution.} was
found.  The bluest wavelength definition of the red pseudo-continuum
which still gives reasonable LOSVD corrections and error statistics for
the index turns out to be 5308.0\,\AA. This is a reduction of $\approx
10$\,\AA\/ in the red-most limit of the original Lick/IDS definition,
equivalent to an increase in the field of view of 13\%.

\begin{figure}
  \includegraphics[width=84mm]{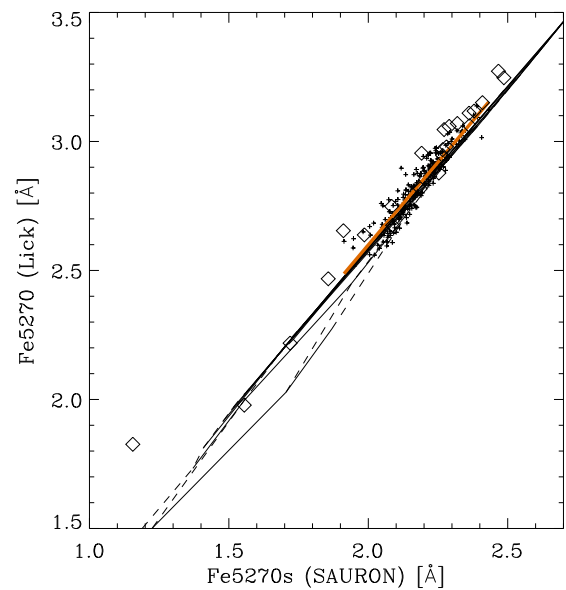}
 \caption{The relation between the new \fes\/ index and the original
   Lick/IDS index Fe5270. The small plus signs represent SAURON
   observations of NGC\,3379 within an eight arcsec radius of the
   center, while the open diamonds show the central measurements of a
   complete sample of early-type galaxies in the Fornax cluster
   \citep{kun00}. The thick solid red line is a linear fit to the
   SAURON data: Fe5270 =~$1.28\, \times$~\fes\/ + 0.03. Overplotted are
   model predictions by \citet{vaz99} at the Lick/IDS resolution.}
 \label{fig:fe5270s}
\end{figure}

The LOSVD correction for a purely Gaussian LOSVD for the new index is
only 5.1\% at 200\,\kms\/ as compared to 10.3\% for the original index.
While the index could be improved in terms of its velocity dispersion
corrections and Poisson statistics it also shows a good linear relation
with the original Lick/IDS index, i.e., it shows a similar dependence
on age and metallicity (see Figure~\ref{fig:fe5270s}). Fortunately, the
re-definition of this index did not significantly change its overall
sensitivity to abundance ratios \citep{kor05}. The best fitting linear
relation determined from stellar population models of \citet{vaz99}
with ages $\ge4$\,Gyr is:

\begin{equation}
  \label{eq:fe5270s}
  \rmn{Fe5270} = 1.26 \times \rmn{Fe5270}_{\rmn{S}} + 0.06 .
\end{equation}
The formal errors of the fit are smaller than $\pm0.01$. In this paper
we use the above equation to convert our \fes\/ measurements to the
Lick Fe5270 index. We note, an empirically determined conversion
formula compares well (Fe5270 =~$1.28\, \times$~\fes\/ + 0.03; see
Figure~\ref{fig:fe5270s}).

In order to maximize the information for the line strength indices we
produce for each galaxy two data-cubes: one containing the \hb, Fe5015
and \mgb\/ indices for the full field of view and one presenting the
new \fes\/ index for a reduced field of view. Due to the adaptive
two-dimensional binning, the final spatial sampling of the two
data-cubes is not identical.

\begin{table*}
  \centering
  \begin{minipage}{140mm}
 \caption{Bandpass definitions of \sauron\/ line strength indices}
 \label{tab:lst_def}
 \begin{tabular}{lccccl}
  \hline
Index     &Blue pseudo-continuum &Central bandpass      &Red pseudo-continuum  & Units  & Source    \\ 
\hline                             
\hb\/   & 4827.875 -- 4847.875 & 4847.875 -- 4876.625 & 4876.625 -- 4891.625 & \AA & \citet{tra98}\\
Fe5015     & 4946.500 -- 4977.750 & 4977.750 -- 5054.000 & 5054.000 -- 5065.250 & \AA & \citet{tra98}\\
Mg\,$b$    & 5142.625 -- 5161.375 & 5160.125 -- 5192.625 & 5191.375 -- 5206.375 & \AA & \citet{tra98}\\
Fe5270     & 5233.150 -- 5248.150 & 5245.650 -- 5285.650 & 5285.650 -- 5318.150 & \AA & \citet{tra98}\\
\fes       & 5233.000 -- 5250.000 & 5256.500 -- 5278.500 & 5285.500 -- 5308.000 & \AA & This paper\\ 
\hline
 \end{tabular}

\end{minipage}

\end{table*}

\subsection[]{Calibration to the Lick/IDS system}
In order to allow a meaningful comparison between stellar population
model predictions and observed data, the index measurements need to be
carefully calibrated to the Lick/IDS system. There are three effects to
account for: (a) the difference in the spectral resolution between the
Lick/IDS system and the \sauron\/ instrumental set-up; (b) the internal
velocity broadening of the observed objects and (c) small systematic
offsets caused by continuum shape differences \citep{worott97,kun00}.

In the following paragraphs we describe the individual steps taken to
ensure an accurate calibration of the \sauron\/ data to the Lick/IDS
system. 


%
\subsubsection[]{Spectral resolution correction}
\label{sec:rescorr}
First we need to adjust for the difference in spectral resolution
between the Lick/IDS system and the \sauron\/ observations for
line strength measurements. The nominal resolution of the Lick/IDS
system in the \sauron\/ wavelength range is $8-9$\,\AA\/ (FWHM) with a
mean of 8.4\,\AA\/ \citep{worott97}. We degrade our spectra to
approximately the Lick/IDS resolution by broadening to an instrumental
resolution of $\sigma = 212$\,\kms.

\subsubsection[]{Line of-sight-velocity-distribution corrections}
\label{sec:veldispcorr}
The observed spectrum of a galaxy is the convolution of the integrated
spectrum of its stellar population(s) by the instrumental broadening
and line of-sight velocity distribution (LOSVD) of the stars. These
effects broaden the spectral features, in general reducing the observed
line strength compared to the intrinsic values. In order to compare the
index measurements for galaxies with model predictions we calibrate the
indices to zero velocity dispersion and the nominal Lick/IDS
resolution.

In the past, most authors have only taken into account the first
moments, $v$ and $\sigma$, of the LOSVD to correct the line strengths
indices.  However, with the availability of high quality measurements
of the higher order terms $h_3$ and $h_4$ which describe non-Gaussian
deviations (see e.g., Paper~III) we need to apply more accurate
corrections. Typically a non-zero $h_3$ term does not affect the Lick
indices significantly, but a non-zero $h_4$ term can have significant
effects \citep{hau98,kun04}.

In order to determine the corrections for each index we use the optimal
template determined for each spectrum individually during the
derivation of the kinematics (for details see Paper~III). Generally,
this optimal template is a good representation of the galaxy spectrum.
For each spectrum a correction factor, $C_{j}(\sigma, h_3, h_4)$, is
determined such that

\begin{equation}
C_{j}(\sigma, h_3, h_4) = I_{j}(\sigma=0, h_3=0, h_4=0) / I_{j}(\sigma, h_3, h_4),
\end{equation}
where $I_{j}$ is the index measured from the optimal template convolved
to the Lick resolution and additionally convolved with the LOSVD given
in brackets. A LOSVD corrected index is then $I_{j}^{corr} =
C_{j}(\sigma, h_3, h_4) \times I_{j}^{raw}$.

\subsubsection[]{Lick/IDS offsets}
\label{sec:offsets}
Although we matched the spectral resolution of the Lick system, small
systematic offsets of the indices introduced by continuum shape
differences are generally present (note that the original Lick/IDS
spectra are not flux calibrated). To establish these offsets we
compared our measurements for stars in common with the Lick/IDS stellar
library \citep{wor94b}. In total we observed 73 different Lick/IDS
stars with 174 repeat observations\footnote{Since the stars have only
  small relative velocities and were typically observed at the center
  of the \sauron\/ field of view we were able to measure the original
  Lick/IDS index Fe5270 for all stars.}.

\begin{figure*}
 \includegraphics[width=120mm]{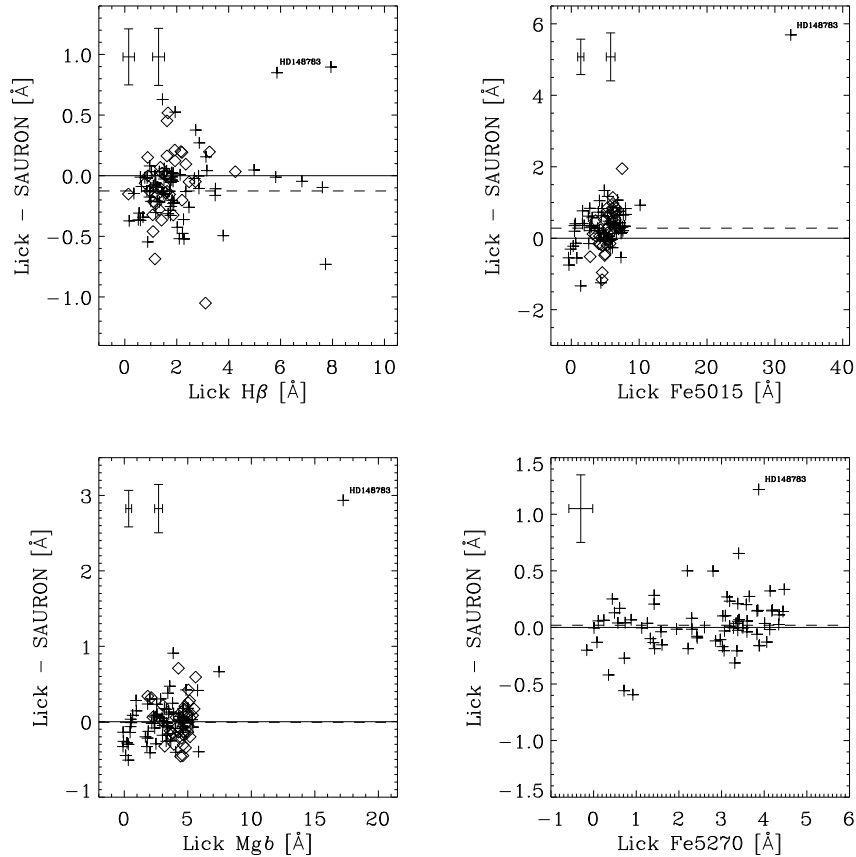}
 \caption{Comparison of Lick/IDS and our measurements for 73 stars in
   common (plus signs). For one star (HD\,148783, spectral type M6III)
   our measurements were drastically different for all four indices.
   The dashed line shows the mean offset derived by a biweight
   estimator (see also Table~\ref{tab:lick_offsets}). Observations of
   galaxies in common between Lick and \sauron\/ are shown as open
   diamonds. A typical error bar for an individual star is shown in the
   upper left corner of each panel. For the galaxy comparisons we plot
   a separate representative error bar. The error is dominated by the
   observational error of the Lick/IDS data. Note that the stellar
   observations span very well the index range covered by the
   galaxies.}
 \label{fig:lick_offsets}
\end{figure*}

Figure~\ref{fig:lick_offsets} shows the difference between Lick/IDS
measurements and ours, where repeat observations of the same star were
averaged. The mean offsets and associated errors for each index are
evaluated by a biweight estimator to minimize the influence of outliers
and are summarized in Table~\ref{tab:lick_offsets}. The formal error in
the offset is taken to be the standard error on the mean $\sigma$ /
$\sqrt{N_{\rmn stars}}$.

As an independent test of this procedure, we compare the Lick/IDS
measurements for the 42 galaxies in common with our survey\footnote{In
  the Lick/IDS survey the \mgb\/, Fe5015, and \hb\/ indices are not
  available for one, three and two galaxies, respectively.}. This is
more difficult than the similar comparison for stars since for
galaxies, e.g., aperture differences and seeing variations can affect
the results. In order to match the Lick/IDS standard aperture of
1\farcs4$\times$4\arcsec\/ we extracted the luminosity-weighted mean of
a circular aperture with radius $r=1\farcs335$ from each line strength
map. It would in principle be better to extract the exact Lick
aperture: this is however not possible since the position angles of the
Lick observations are unknown to us. For this comparison the \sauron\/
measurements were not corrected for emission but internal line
broadening was taken into account (see Section~\ref{sec:veldispcorr}).

The results are overplotted in Figure~\ref{fig:lick_offsets} as open
diamonds and the offsets, evaluated with a biweight estimator, are
listed in Table~\ref{tab:lick_offsets}. The overall offsets inferred
from the galaxies are in good agreement with the offsets derived from
the stars. To further test the offsets for Lick indices derived from
flux calibrated data we made use of the \citet{jon97} library. First
the data was broadened to the spectral resolution of the Lick/IDS
system with a wavelength dependent Gaussian assuming a constant
spectral resolution of FWHM = 1.8\,\AA\/ for the Jones stars. Then we
compared the index measurements for the 128 stars in common between the
Jones library and the Lick observations \citep{wor94b}. The offsets and
associated errors, derived with a biweight estimator, are listed in
Table~\ref{tab:lick_offsets}, column 4 \citep[see also,][]{nor05} and
are in good agreement with the offsets derived from the \sauron\/
observations for the \hb\/ and Fe5015 indices. The agreement for the
\mgb\/ and Fe5270 indices is less good, but differences are still small
with offsets $<0.1\,$\AA.  Furthermore, our determination of Lick
offsets derived from Jones stars is in excellent agreement with an
earlier investigation carried out by \citet[][Table 9]{worott97}.
Although the offsets are small, they can significantly change the age
and metallicity estimates.  Particularly, the \hb\/ index presents a
problem, since one can find with the currently available stellar
population models age estimates older than the age of the universe if
offsets are applied. In the end we decided to apply Lick offsets only
for the \hb\/ and Fe5015 indices (see Table~\ref{tab:lick_offsets},
column 5) since all three offset determination methods give consistent
results. For the \mgb\/ and Fe5270 indices no offsets are applied since
the \sauron\/ stellar observations indicate offsets consistent with
zero. For all indices we quote conservative offset errors which reflect
the information derived from the stars and galaxies.

\begin{table*}
  \centering
  \begin{minipage}{110mm}
 \caption{Offset estimates to the Lick/IDS system}
 \label{tab:lick_offsets}
 \begin{tabular}{lcccc} \hline
Index  & Offset (stars)        &  Offset (galaxies)    &  Offset (Jones stars) & Applied offsets    \\ 
 (1) & (2) & (3) & (4) & (5) \\
\hline               
\hb\/  & $-0.13\pm0.03$\,\AA   &  $-0.06\pm0.04$\,\AA  &  $-0.12\pm0.02$\,\AA  & $-0.13\pm0.05$\,\AA\\
Fe5015 & $+0.28\pm0.05$\,\AA   &  $+0.27\pm0.09$\,\AA  &  $+0.23\pm0.04$\,\AA  & $+0.28\pm0.05$\,\AA\\
\mgb   & $-0.01\pm0.03$\,\AA   &  $+0.01\pm0.04$\,\AA  &  $-0.08\pm0.02$\,\AA  & $+0.00\pm0.05$\,\AA\\
Fe5270 & $+0.02\pm0.02$\,\AA   &  --                   &  $-0.07\pm0.02$\,\AA  & $+0.00\pm0.05$\,\AA\\
\hline
 \end{tabular}

\medskip

Notes: Column~(1) gives the index name, while column~(2) and~(3) give
the mean offset (Lick - \sauron) to the Lick/IDS system evaluated from
the stars and galaxies in common, respectively. Column~(4) shows the
offsets derived from the comparison of the \citet{jon97} library to the
Lick/IDS observations for stars in common. Column~(5) shows the final
offsets we applied to our data and their mean adopted error.
\end{minipage}

\end{table*}

\subsection[]{Estimating errors for line strength indices}
\label{sec:lst_errors}
Several authors have evaluated the sources of errors for line strength
indices \cite[][and references therein]{card98}. They can be divided
into random and systematic errors. The treatment of systematic errors
is generally very difficult, while the random errors can be determined
quite accurately. In the following paragraphs we concentrate on the
random errors.

The main ingredient to evaluate the random errors for line strength
indices in the \sauron\/ system are the noise spectra which are
provided by the data reduction procedure for each individual lenslet
(see Paper~I). The noise spectrum represents the errors propagated
through the full data-reduction until the final, merged and binned
data-cube from which velocities, velocity dispersions, line strength
indices etc. can be determined.

Before measuring a given index on a spectrum one needs to know: a) the
average recession velocity in order to place the index band-passes
correctly and b) a measure of the LOSVD in order to correct the index
for broadening effects (see Section~\ref{sec:veldispcorr}). Both
parameters have associated uncertainties (for details see Paper~III)
which will contribute to the final index error.

For each galaxy, the random error for each index measurement was
estimated via a Monte Carlo approach. For every spatial bin, N
realizations of the associated spectrum were produced based on its
corresponding noise distribution (assumed to be Gaussian). Likewise, N
values of the recession velocity and velocity dispersion were chosen
within the uncertainties of their measured values, thus creating a
total number of N$^3$ realizations of the measured index, from which
the rms uncertainty was determined. A typical number for N in our
analysis was 30. For bins where emission in \hb, \oiii\/ or \nishort\/
was detected and subtracted we add the associated errors in quadrature
(see Section~\ref{sec:emission_corr} and Paper~V).

Concerning the systematic errors we consider two main contributions:
(i) errors from the continuum correction for the \hb\/ line strengths
and (ii) the error in the calibration to the Lick system. The continuum
correction errors (see Section~\ref{sec:cont}, assumed to be at a
constant level of 0.1\,\AA) are added to the affected bins while the
Lick system offsets are quoted in Table~\ref{tab:lick_offsets} and are
not added to the individual data points, but shown as global offset
errors in figures.

\section{Observed Line strength Maps}
\label{sec:maps}
Figures~\ref{fig:maps1}-\ref{fig:maps12} below present maps of the
absorption line strengths of the 48 objects, ordered by increasing NGC
number. For each galaxy we show the total intensity reconstructed from
the full wavelength range of the \sauron\ spectra (see also Paper~III),
and the two-dimensional line strength distributions of \mgb, Fe5015,
\fes\/ and \hb, overplotted with isophotes of the reconstructed image
spaced by single magnitude steps. The maps are all plotted with the
same spatial scale, and oriented with respect to the \sauron\ field for
presentation purposes. The relative directions of north and east are
indicated by the orientation arrow next to the galaxy title (we note
that the orientation and binning of the maps is identical to Papers~III
and V).  The maximum and minimum of the plotting range is given in the
tab attached to each parameter map, and the color bar indicates the
color table used. In order to allow for an easy comparison between
galaxies the plotting range of the \hb\/ maps is fixed to 1.0 --
2.9\,\AA\/ (with the exception of NGC\,3032, 3156, 4150 which show
larger \hb\/ absorption strengths in the central parts). The colors are
adjusted such that blue to green shades correspond to the \hb\/
strength predicted for old stellar populations ($\sim$12\,Gyr), while
stronger \hb, corresponding to the presence of younger stellar
populations, are represented by red and yellow shades \citep[see
e.g.,][]{TMB03}. For the metal line maps we use an independent plotting
range for each galaxy in order to better visualize the line strengths
gradients across the maps.

For some galaxies we found non-physically large values of the Fe5015
line strength in the outer regions where many individual lenslets are
averaged to achieve the target S/N of 60. The same effect can be seen
to a much lesser extent in \mgb\/ maps. Similar to the continuum
variations which cause some of the \hb\/ measurements to be corrupted
in the outskirts, we suspect that the \mgb\/ and Fe5015 indices are
also affected (see Section~\ref{sec:cont}). Due to the degeneracy
between abundance ratio variations and continuum effects we cannot
apply a continuum correction for the metal lines. Therefore, we decided
to remove the most affected bins by hand from the final data-cubes.
These bins are indicated with grey color in the line strength maps
presented in Figures~\ref{fig:maps1}-\ref{fig:maps12}.

\subsection{Overview of the line strength maps}
The maps in Figures~\ref{fig:maps1}-\ref{fig:maps12} show a wealth of
structures and we give comments on individual galaxies in
Appendix~\ref{sec:list}. Some general trends are apparent and we
discuss these briefly in the following. A full stellar population
analysis of the line strength maps with the help of stellar population
models will be presented in a future paper in this series.

The metal line strength maps often show negative gradients with
increasing radius roughly consistent with the morphology of the light
profiles. Remarkable deviations from this trend exist: NGC\,3032,
3156, 4150, and 4382 show a central depression in \mgb\/ line strength.
This structure is always accompanied by a strong peak in the \hb\/ maps
indicating the presence of a recent post-starburst or even low level
ongoing star-formation. Interestingly, the Fe maps of post-starburst
galaxies do not always show a depression such as the \mgb\/ index which
can be understood in the reduced sensitivity to stellar population age
of Fe indices.

Significant, but low-level, deviations between isoindex contours of the
metal indices and the isophotes are discussed separately in
Section~\ref{sec:mgb_cont}. We find that enhanced \mgb\/ absorption
strength is connected to fast-rotating components \citep[as defined
in][hereafter Paper~IV]{cap05a} in the galaxies. Perhaps the strongest
case in our sample is presented in the \mgb\/ map of NGC\,4570 where
the \mgb\/ strength along the major axis runs through a dip outside the
bulge region and then begins to rise towards larger radii where the
galaxy kinematics start to be dominated by an outer disk \citep[see
also ][for a long-slit study of S0s]{fis96}.

For galaxies with weak \hb\/ absorption ($\simeq 1.4$\, \AA) the maps
show typically mild positive gradients or are consistent with being
flat. A few galaxies (e.g., NGC\,3489, NGC\,7332 and NGC\,7457) show
\hb\/ maps with an overall elevated \hb\/ line strength indicating a
spatially extended, recent star-formation episode. Other galaxies
(NGC\,4382, NGC\,4459 and NGC\,4526) exhibit a more centrally
concentrated region of relatively strong \hb\/ absorption pointing
towards a more localized star-formation event.

Regions of strong \hb\/ absorption are often associated with dust
features and detected in emission (see emission maps and unsharp-masked
images in Paper~V). The case of NGC\,4526 deserves a special note,
since here one can see the direct connection between an almost perfect
dusty disc seen close to edge on (apparent in the reconstructed image),
and the location of strong \hb\/ absorption indicating the presence of
young ($\sim$1\,Gyr) stars.

\subsection{\mgb\/ isoindex contours versus isophotes}
\label{sec:mgb_cont}
One of the most interesting aspects of integral-field spectroscopy is
the capability to identify two-dimensional structures. For the first
time we can use this in connection with line strength indices and
compare isoindex contours with the isophotal shape. One might expect
the index to follow the light in slowly-rotating giant elliptical
galaxies, since the stars are dynamically well mixed. However, one
could imagine that dynamical substructures with significantly different
stellar populations could leave a signature in line strength maps which
are sensitive to e.g., metallicity.

\begin{figure}
 \includegraphics[width=84mm]{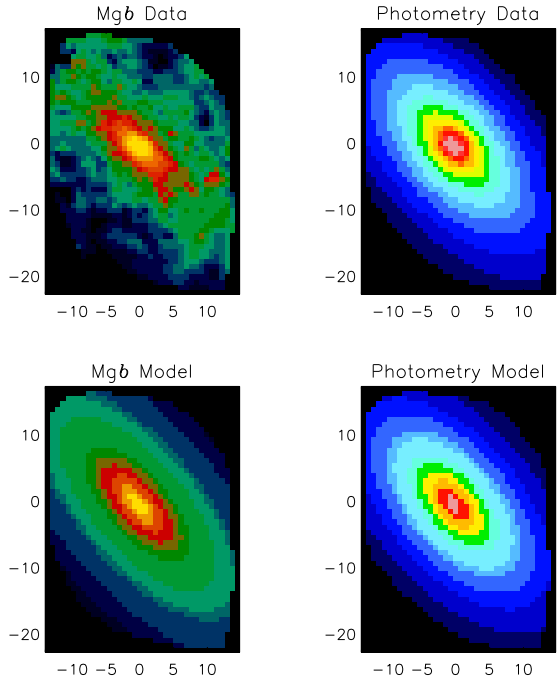}
 \caption{{\bf Top row:} The interpolated \mgb\/ map and the
   reconstructed image of NGC\,3377. {\bf Bottom row:} The best fitting
   elliptical model with constant position angle and ellipticity to the
   \mgb\/ map and the reconstructed image.  The best fitting
   ellipticity for the isophotes and the \mgb\/ map is $0.473\pm0.003$
   and $0.573\pm0.022$, respectively. The x- and y-axis are given in
   arcsec; North is up and East to the left.}
 \label{fig:ls_shape_example}
\end{figure}

The \mgb\/ index is the best determined index in our survey and
potential differences between isoindex contours and isophotes should be
most apparent. Indeed we find a number of galaxies where the isoindex
contours clearly do not follow the isophotes, e.g., NGC\,4570 and
NGC\,4660. Both galaxies show fast-rotating components (see Paper~III)
along the direction of enhanced \mgb\/ strength. In order to further
study such a possible connection we determine the best fitting
($\chi^2$ sense) simple, elliptical model for each of the reconstructed
images and \mgb\/ maps in our survey. For this model we impose a
constant ellipticity and position angle as function of radius, where
the position angle is determined as the average position angle of the
reconstructed image. An example of this procedure is shown in
Figure~\ref{fig:ls_shape_example}. We then compare the derived
characteristic ellipticities for the isophotes and \mgb\/ maps with
each other. Clearly such a simple model does not give a good fit to all
galaxies, but it provides us with a robust, global shape measurement of
the isoindex contours as compared to the isophotes.

In order to avoid confusion by the presence of significant dust
absorption we exclude the following galaxies from the analysis:
NGC\,2685, NGC\,3032, NGC\,3156, NGC\,3489, NGC\,4374, and NGC\,4526.
Furthermore we exclude NGC\,4486 because of the presence of the
non-thermal emission from the well-known jet. This leaves a sample of
41 galaxies.

\begin{figure}
  \includegraphics[width=84mm]{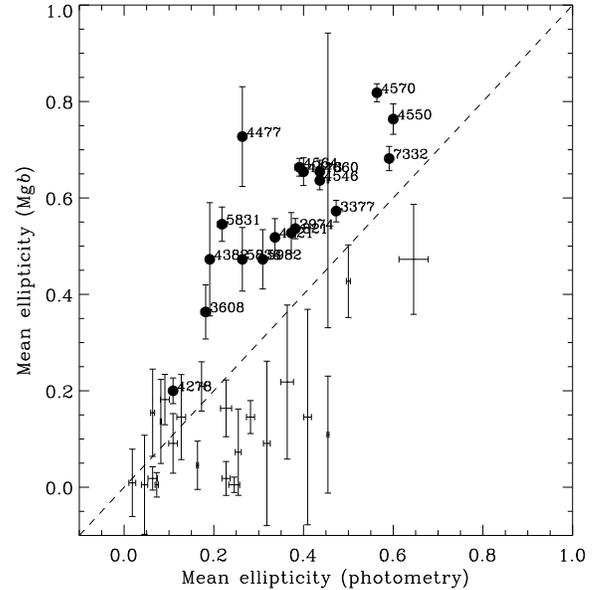}
 \caption{Comparison of the average ellipticity of constant \mgb\/
   strength with the best fitting elliptical model of the isophotes.
   Errors are evaluated by a Monte-Carlo simulation and represent $1
   \sigma$ errors. All galaxies which are more than $2\sigma$ above the
   one-to-one line are indicated by filled circles and their galaxy
   names.}
 \label{fig:ls_shape}
\end{figure}

The results of the ellipse fitting are shown in
Figure~\ref{fig:ls_shape}. Errors are evaluated via 50 Monte Carlo
simulations per map where the \mgb\/ strength of each bin is varied
according to the index errors. Additionally, we allowed for a 5\%
fraction of the data values to be considered as outliers and varied
those data points by 5 times its original error. We required a minimum
error of 0.08\,\AA\/ per bin\footnote{In the central, high S/N regions
  of the \mgb\/ maps our internal error estimates are very small and we
  use a minimum error of 0.08\,\AA\/ to reflect systematic errors.} and
also excluded all bins covering more than 20 individual lenses from the
analysis.

Applying a $2\sigma$ error cut we find 16 out of the 41 galaxies to be
consistent with no deviation between isophotes and isoindex contours.
Seven galaxies appear to have rounder isoindex contours than the
isophotal shape (NGC\,2695, 2699, 2768, 3384, 3414, 4262, and 5845).
However, 18 galaxies appear to have more flattened \mgb\/ contours than
the isophotes (NGC\,821, 2974, 3377, 3608, 4278, 4382, 4473, 4477,
4546, 4550, 4564, 4570, 4621, 4660, 5831, 5838, 5982, and 7332). Most
of these galaxies show a high degree of rotational support and
significant $h_3$ terms. Thus, the flattened \mgb\/ distribution
suggests that the fast-rotating components in these galaxies exhibit a
stellar population different from the main body. The enhanced \mgb\/
strength can be interpreted to first order as higher metallicity and/or
increased [Mg/Fe] ratio. Furthermore, it is evident from the
line strength maps that the detailed isoindex shapes cannot always be
modeled with a simple elliptical structure. For example, in NGC\,4570
one can see a central bulge component with a strong radial \mgb\/
gradient while at larger radii the onset of a disk component with
strong \mgb\/ is visible \citep[e.g.,][]{bosch98}. A detailed analysis
of the line strength distribution and its possible connection to the
kinematics will be presented in a forthcoming paper.

%

\section{Line strength gradients and aperture corrections}
\label{sec:gra_ap}
\subsection{Line strength gradients}
\label{sec:ls_grad}
A number of previous studies have investigated line strength gradients
with the help of long-slit data
\citep[e.g.,][]{gor90,car93,dav93,fis96,meh03}. Line strength gradients
can be used to study the formation history of early-type galaxies since
different formation models predict different gradients.  In a nutshell,
monolithic collapse models \citep{car84} predict steep metallicity
gradients with metal rich centers whereas hierarchical models,
following a merger tree, predict shallower gradients due to the
dilution of any line strength gradients existing in the pre-merger
units \citep{whi80}.

We now derive line strength gradients from the \sauron\/ sample. First,
all indices are expressed in magnitudes (as is the well known Mg$_2$
index) which is indicated by a prime sign [$^\prime$]. The conversion
between an index measured in \AA\/ and magnitudes is

\begin{equation}
  \label{equ:mgbp}
  {\rmn{index}}\,^\prime= -2.5 \log \left( 1 - \frac{\rmn{index}}{\Delta \lambda}\right),
\end{equation}
where $\Delta \lambda$ is the width of the index bandpass.

In Section~\ref{sec:mgb_cont}, we have mentioned the fact that overall,
the morphology of the line strength maps resembles the corresponding
photometry. This motivated our choice to derive robust and simple line
strength gradients by averaging the indices along lines of constant
surface brightness (isophotes) with equal steps in $\log$~flux. The
average line strength in each radial bin is derived after applying a
$3\sigma$ clipping algorithm. The radius for each bin is calculated as
the median major axis radius normalized to the effective radius along
the major axis $a_{\rmn e} = R_{\rmn e}/\sqrt{1-\epsilon}$, where
$R_{\rmn e}$ is the effective radius and $\epsilon$ is the average
ellipticity of the galaxy within the \sauron\/ field. The effective
radii $R_{\rmn e}$ (see Table~\ref{tab:line_re}) are derived with a
$R^{1/4}$ growth curve analysis from our wide-field MDM (1.3m) imaging
survey of the E/S0 sample, supplemented by archival HST images. The
photometry and analysis thereof will be presented in a forthcoming
paper of this series.

\begin{figure}
  \includegraphics[width=84mm]{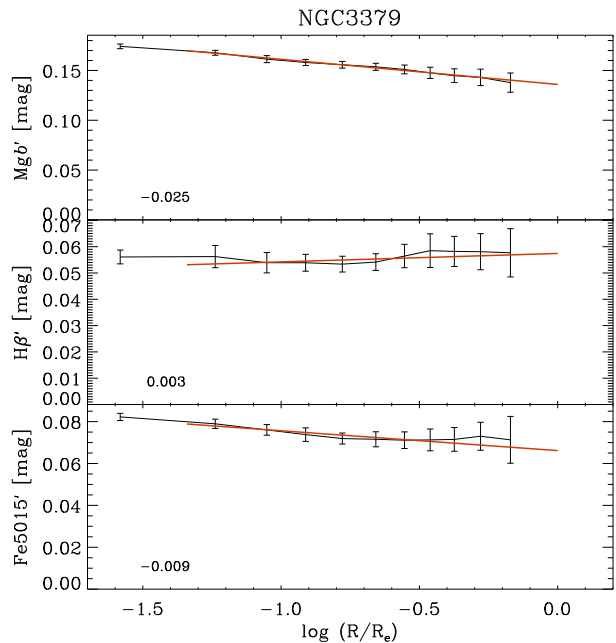}
  \caption{Line strength gradients for NGC\,3379 as function of $\log\,
    (R/R_{\rmn e})$ along the major axis. Error bars reflect the
    $1\sigma$ scatter of data-points along a given band of constant
    surface brightness. The red line represents the best-fitting
    straight line to the data. The radial extent of the red line
    reflects the fitting range in $\log R/R_{\rmn e}$. The fitted line
    strength gradients are noted in the lower left of each panel.}
  \label{fig:ls_grad}
\end{figure}

For each galaxy we fit an error-weighted straight line to the index
values at a given radius such that gradients are defined as

\begin{equation}
\Delta\, {\rmn{index}}\,^\prime = \frac{\delta\,{\rmn{index}}\,^\prime}{\delta\, \log (R/R_{\rmn e})}.
\end{equation}
The error of the index value in each surface brightness bin is taken as
the $1\sigma$ scatter of the data-points within this bin after applying
a $3\sigma$ clipping algorithm. We further restrict the fitting range
to radii between 2\arcsec\/ and $R_{\rmn e}$. An example fit to the
data of NGC\,3379 is shown in Figure~\ref{fig:ls_grad}.

\begin{figure*}
  \includegraphics[width=84mm]{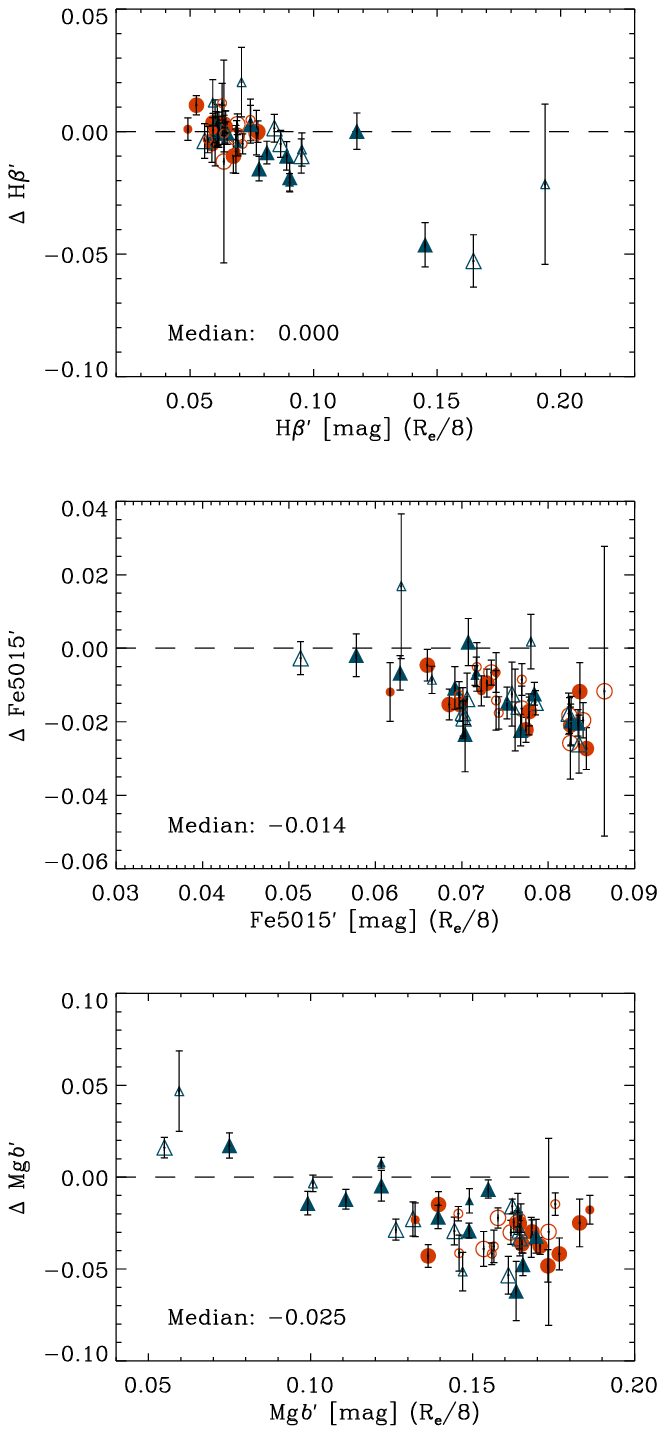}
  \includegraphics[width=84mm]{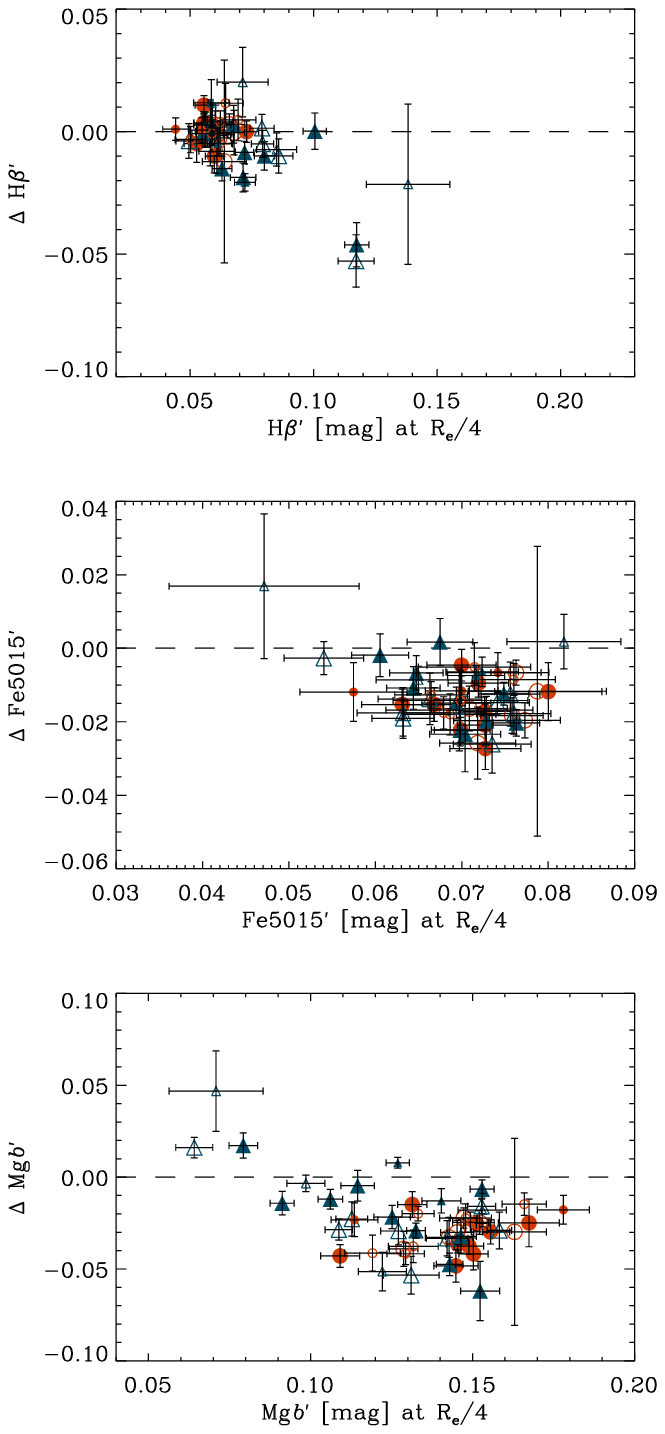}
  \caption{{\bf Left column of panels:} Index gradients as function of 
    luminosity-weighted index values within a circular aperture of one
    eights of an effective radius \re. The red filled and open circles
    are cluster and field ellipticals, respectively; blue filled and
    open triangles are cluster and field S0s, respectively. The
    smaller symbols indicate galaxies for which our data cover less
    than half the effective radius. Error bars are formal errors from
    the straight line fits. A gradient of 0.0, i.e. a flat relation, is
    indicated by the dashed lines. Median gradients are given in the
    lower left of each panel.  See text for details. {\bf Right column
      of panels:} Index gradients as function of (fitted) line strength
    at 1/4 effective radius along the major axis. Symbols are the same
    in both diagrams.}
  \label{fig:ls_grad_sum1}
\end{figure*}

In Figure~\ref{fig:ls_grad_sum1} we present for the full sample the
fitted gradients for the \hb, Fe5015, and \mgb\/ indices plotted
against the average (luminosity-weighted) index values within a
circular aperture of one eights of an effective radius ($R_{\rm e}/8$,
see Section~\ref{sec:central} for details). For elliptical galaxies the
\hb\/ gradients have a mean value of $0.000\pm0.006$, i.e., they are
consistent with a flat relation \citep[see also][]{meh03}. The
lenticular galaxies show a range in \hb\/ gradients where stronger
(negative) gradients are correlated with increasing central \re\/ \hb\/
absorption strengths. This is mostly a consequence of the presence of
galaxies in our sample which harbor young stars in the central regions,
but have little or no star formation in the outer parts. If these young
stars amount only to a small fraction in mass, the gradients will
within a few Gyrs become shallower and the central \hb\/ absorption
strength will decrease, moving the galaxies back to the bulk of points
in Figure~\ref{fig:ls_grad_sum1}.

We find mostly negative gradients for the metal line gradients,
consistent with the literature \citep[e.g.,][]{dav93,car93,kob99}.
Median gradients for Fe5015$^\prime$ and \mgbp\/ are -0.014 and -0.025,
respectively. Using the full sample we find significant correlations
between the line strength gradients and the average values within \re\/
for Fe5015 and \mgb. The correlation coefficient derived from a
(non-parametric) Spearman rank-order test is $-0.55 (<1\%)$ and $-0.49
(<1\%)$, for the Fe5015$^\prime$ and \mgbp\/ indices, respectively. The
probability that the parameters are not correlated is given in
parentheses.

A possible correlation between the Mg gradient (as measured by the
Mg$_2$ index) and the central Mg line strength had been suggested by
\citet[][see also Carollo et al. 1993]{gon95}, but some other
investigations failed to detect a significant correlation
\citep[e.g.,][]{kob99,meh03}. New long-slit observations of a sample of
82 early-type galaxies presented by \cite{san04} do confirm the trend
found by \citet{gon95}.

\begin{figure}
  \includegraphics[width=84mm]{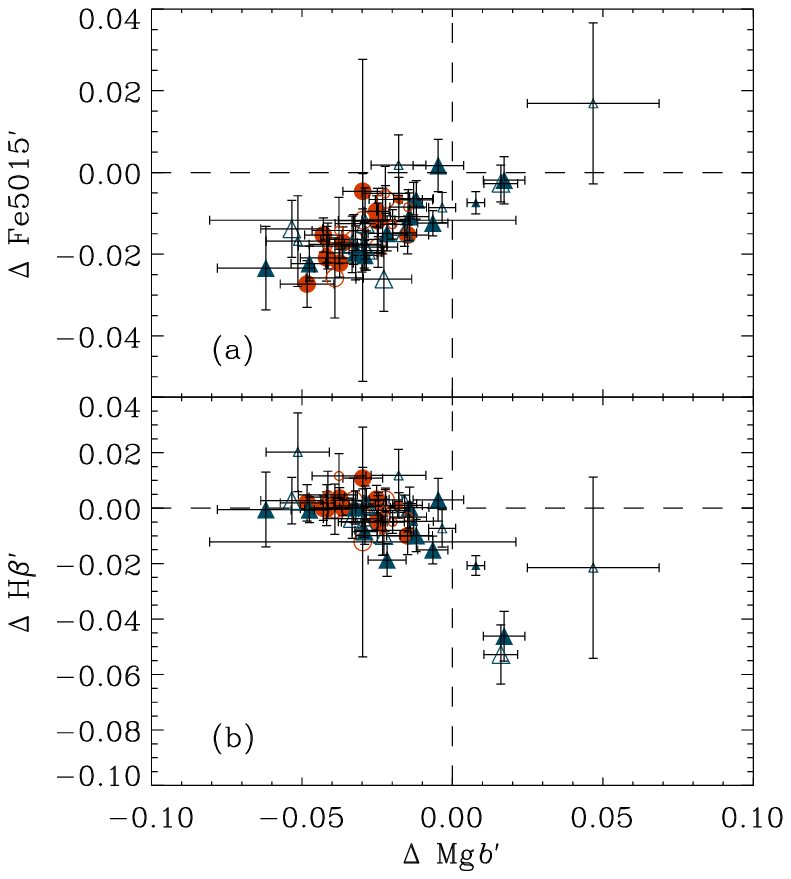}
  \caption{(a) Fe5015$^\prime$ gradients {\em versus}\/ \mgb$^\prime$
    gradients. (b) \hb$^\prime$ gradients {\em versus}\/ \mgb$^\prime$
    gradients. The red filled and open circles are cluster and field
    ellipticals, respectively; blue filled and open triangles are
    cluster and field S0s, respectively. The smaller symbols indicate
    galaxies for which we do not cover half the effective radius.
    Error bars are formal errors from the straight line fits.}
  \label{fig:grad_grad}
\end{figure}

In our sample the correlation for the \mgbp\/ index is largely driven
by the (lenticular) galaxies with \mgbp\/ $< 0.13$. These galaxies are
the ones with young stars in the central regions. This can be seen in
Figure~\ref{fig:grad_grad}b where we show the relation between \hbp\/
and \mgbp\/ gradients. There is also a good correlation between the
gradients of the \mgbp\/ and Fe5015$^\prime$ indices as shown in
Figure~\ref{fig:grad_grad}a \citep[see also][]{fis96}. This indicates
that both indices are sensitive to similar stellar population
parameters (e.g., metallicity).

In summary it appears that in our sample there are significant
correlations of the line strength gradients with the central
line strength in early-type galaxies. The relations are predominantly
driven by the lenticular galaxies and furthermore by the presence of
young stellar population in the central regions of galaxies.
Environmental differences do not seem to play a major role. It is now
interesting to explore if the above correlations also hold if
line strength gradients are compared to index strength at larger radii,
i.e. more representative regions for the galaxies as a whole.

In Figure~\ref{fig:ls_grad_sum1} (right hand side panels) we plot the
fitted gradients for the \hb, Fe5015, and \mgb\/ indices against the
index value {\em at}\/ 1/4 effective radius along the major axis. The
index value at 1/4 effective radius is taken to be the value predicted
by the fit. We have chosen this particular reference point, since it is
within the range of the observed line strength gradients for all
galaxies and is more representative of the galaxy as a whole rather
than being biased to the central regions.

In this new diagram we do not find a significant correlation for the
Fe5015$^\prime$ index. For the \hb\/ index there is still a correlation
visible but it is exclusively driven by the three galaxies in the lower
right part of the plot. The situation for \mgbp\/ appears more complex.
There is no significant correlation for the full sample while the S0s
show a trend similar to the one derived for the 1/8 effective radius
values. The elliptical galaxies even appear to follow a {\em
  positive}\/ correlation between line strength gradient and \mgbp\/
strength at $R_{\rmn e}/4$. However, the latter trend is not
significant.

We conclude that for early-type galaxies in our sample there is a
correlation between line strength gradients of \hb, Fe5015 \& \mgb\/
and centrally averaged line strength values \re. These correlations are
driven to a large extent by the presence of young stellar populations
in the central regions of (typically lenticular) early-type galaxies
and thus are caused by galaxies which have recently experienced
secondary star-formation which may account only for a minor fraction of
the total galaxy mass.

\subsection{Aperture corrections}
\label{sec:ls_aper}
For studies of galaxies over a significant redshift and size range
aperture corrections have to be applied in order to allow for a fair
comparison of kinematic and stellar population parameters
\citep[e.g.,][]{jor95}. In Paper~IV we present an aperture correction
for the luminosity weighted second moment of the line of-sight velocity
distribution ($\sigma$) while here we present aperture corrections for
the line strength indices \hb, Fe5015, and \mgb.

\begin{figure}
  \includegraphics[width=84mm]{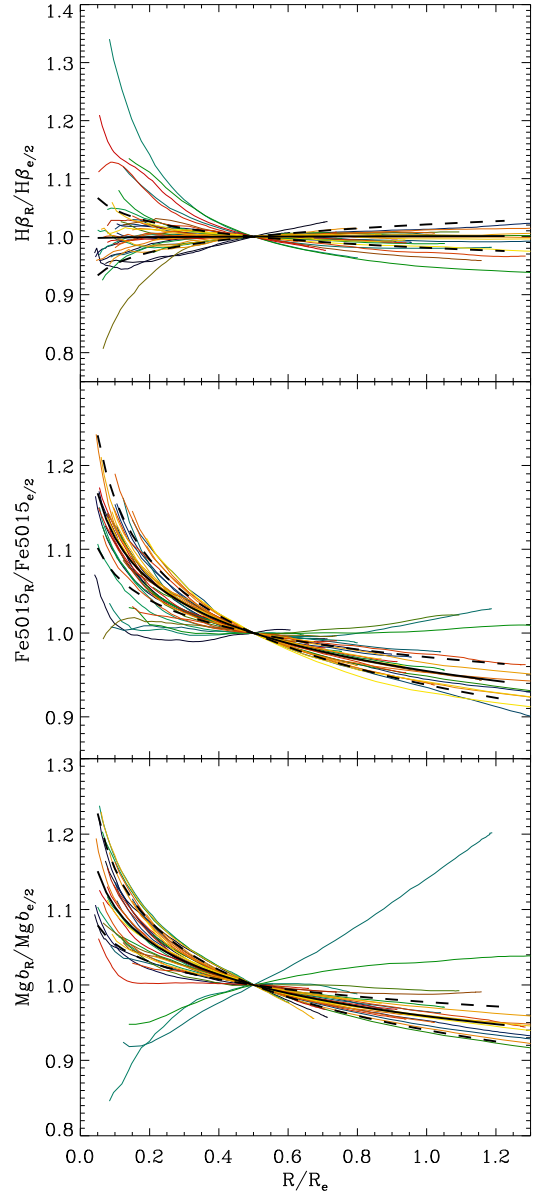}
  \caption{Luminosity-weighted index measurements within an aperture of
    radius $R$, normalized to its value at $1/2~R_{\rmn e}$. Each
    galaxy with data out to at least $1/2~R_{\rmn e}$ is shown with the
    thin colored lines. Data points inside $R=2\arcsec$ are excluded to
    minimize seeing effects. The black line is the median power-law
    relation $(index_{\rmn R}/index_\rmn{norm}) = (R/R_\rmn{norm})^{p}$
    where $p$ is given in Table~\ref{tab:aper}. The robustly estimated
    $1\sigma$ errors of the power law fits are indicated by the dashed
    black lines. See text for details. The three galaxies with
    $(\rmn{Mg}\,b_{\rmn R} / \rmn{Mg}\,b_{\rmn e/2}) < 1.0$ at small
    radii are NGC\,3032, NGC\,3156 and NGC\,4150.}
  \label{fig:aper}
\end{figure}

For this we show in Figure~\ref{fig:aper} the luminosity-weighted
line strength measurements within an aperture of radius $R$, normalized
to their value at half of the effective radius. The profiles were
measured by co-adding the \sauron\/ spectra within circular apertures
of increasing radius for all 39 early-type galaxies which have data out
to at least half of the effective radius. If the \sauron\/ FoV does not
cover the complete circular aperture at a given radius, we calculate an
equivalent radius of a circular aperture which matches the area of the
\sauron\/ FoV.

The metal line indices show generally a decreasing profile with
increasing $(R/R_{\rmn e})$, while the \hb\/ index shows a wide range
in aperture corrections. Typical aperture corrections are small where a
correction from 1/4 effective radius to one effective radius would
amount to approximately 0.5\% for the \hb\/ index and $\sim$9\% for the
metal indices. However, it is clear that substantial galaxy-to-galaxy
variations in the profiles exist. Particularly, for galaxies harboring
a young stellar population in the center (i.e.  positive \mgb\/
gradients) the application of an average aperture correction can be
severely wrong.

The best fitting power-law relations to the line strength measurements
as function of normalized radius $(\frac{R}{0.5~R_{\rmn e}})$ were
determined for each galaxy individually by fitting all available data
outside $R = 2\arcsec$. Then a median power-law index and a robustly
estimated scatter was derived. These median power-law indices for each
absorption feature are only provided as a convenient approximation of
the true aperture corrections and presented in Table~\ref{tab:aper}.
They are in reasonable agreement with previously published aperture
corrections \citep{jor95,meh03}.

\begin{table}
  \centering
  \begin{minipage}{71mm}
    \caption{Aperture corrections for line strength indices}
    \label{tab:aper}
    \begin{tabular}{rll}
      \hline 
$ (\rmn{H}\beta_{\rmn R} / \rmn{H}\beta _\rmn{norm})$& $=$& $(R/R_\rmn{norm})^{+0.001\pm0.029}$\\
$  (\rmn{Fe5015}_{\rmn R} / \rmn{Fe5015}_\rmn{norm})$& $=$& $(R/R_\rmn{norm})^{-0.067\pm0.025}$\\
$  (\rmn{Mg}\,b_{\rmn R} / \rmn{Mg}\,b_\rmn{norm})$  & $=$& $(R/R_\rmn{norm})^{-0.061\pm0.028}$\\
\hline
\end{tabular}
\end{minipage}
\end{table}

\section{Integrated index--$\sigma$ relations}
\label{sec:central}
After the presentation of the complete line strength maps
(Section~\ref{sec:maps}) and the average gradients
(Section~\ref{sec:gra_ap}) we now present luminosity-weighted
line strength indices averaged over a circular aperture. We focus on
one key diagnostic: the index--$\sigma$ relations.

Our integrated measurements were derived by averaging the
(luminosity-weighted) spectra within a circular aperture with a radius
of \re, but imposing a minimum aperture of $2\farcs4 \times 2\farcs4$
($3 \times 3$ pixel). From this high S/N spectrum we re-measure the
kinematics (V, $\sigma$, $h_3$ and $h_4$) as described in Paper~III
\cite[see also][]{cap04} and then determine the line strength as
presented in Section~\ref{sec:indices}. Because of limited field
coverage we do not quote central averaged \fes\/ measurements for
eleven galaxies (NGC\,474, 524, 821, 2695, 2768, 3608, 4270, 5198,
5308, 5846, and 5982). In a further eight galaxies the \re\/ aperture
is not fully covered by our maps, however, only a few bins,
corresponding to less than 10\% coverage, are missing and we judge the
average spectrum to be still a good representation of the galaxy.

In order to provide a more global measurement for each galaxy we also
derived a central averaged spectrum from all data available within one
effective radius. Because of the sometimes severely limited field
coverage of the \fes\/ index we do not quote one effective radius
averages for it.  Furthermore, since our line strength maps of the
other indices do not always cover the full area of one effective radius
we need to asses the degree of necessary aperture corrections. The
galaxies with the smallest coverage (NGC\,4486 and NGC5846) feature
line strength data out to about 30\% of $R_{\rmn e}$. Even in these
extreme cases we estimate corrections to be small ($<8\%$; see
Section~\ref{sec:ls_aper}). For the 32 galaxies for which the data
extends to less than one effective radius, we apply the corrections of
the form given in Table~\ref{tab:aper}. The median coverage of the line
strength maps is $0.8 R_{\rmn e}$. A subset of one $R_{\rmn e}$ index
measurements is used in Paper~IV to determine stellar $M/L$ ratios and
compare them with high fidelity dynamical estimates of the global $M/L$
inside one $R_{\rm e}$.

The final, luminosity-weighted line strength measurements within a
circular aperture of $(R_{\rmn e}/8)$, and within $R_{\rmn e}$, are
given in Table~\ref{tab:line_re}. The formal line strength errors are
smaller than $\pm0.01$\,\AA\/ for all galaxies. We note, however, that
there are systematic errors which we estimate to be of the order of
0.06, 0.15, 0.08, 0.06\,\AA\/ for the \hb, Fe5015, \mgb, and \fes\/
indices, respectively. For the luminosity weighted averages of the
velocity dispersion we adopt an error of 5\% (see Paper~IV).


\begin{table*}
\begin{center}
  \begin{minipage}{135mm}
 \caption{List of line strength measurements within a circular aperture of $R_{\rm e}/8$ and $R_{\rm e}$}
 \label{tab:line_re}
 \begin{tabular}{lrcrccccrcc}
  \hline
Name & $R_{\rmn e}$ & $R_{\rmn max}/R_{\rmn e}$ & $\sigma_{{\rmn e}/8}$  & \hb  &Fe5015 & \mgb & \fes  & \hb  & Fe5015  & \mgb \\
     & [\arcsec] & & [\kms] & [\AA] & [\AA]  & [\AA] & [\AA] & [\AA] & [\AA]& [\AA]\\
     & & &\reb  & \reb &  \reb & \reb & \reb &  $R_{\rmn e}$ & $R_{\rmn e}$ & $R_{\rmn e}$\\
 (1) & (2) & (3) & (4) & (5) & (6) & (7) & (8) & (9) & (10) & (11)\\
\hline
%
%
  NGC474 &  29 & 0.71 & 163 &  1.68 &  5.45 &  4.11 &  --   &  1.81 &  4.85 &  3.45 \\
  NGC524 &  51 & 0.61 & 249 &  1.39 &  5.57 &  4.55 &  --   &  1.50 &  5.38 &  4.20 \\
  NGC821 &  39 & 0.62 & 202 &  1.53 &  5.32 &  4.35 &  --   &  1.57 &  4.65 &  3.72 \\
 NGC1023 &  48 & 0.56 & 206 &  1.49 &  5.60 &  4.51 &  2.34 &  1.51 &  4.98 &  4.19 \\
 NGC2549 &  20 & 0.90 & 142 &  2.07 &  5.85 &  4.05 &  2.40 &  2.02 &  5.10 &  3.54 \\
 NGC2685 &  20 & 1.34 &  83 &  2.01 &  5.05 &  3.57 &  2.19 &  2.04 &  4.21 &  3.04 \\
 NGC2695 &  21 & 0.95 & 223 &  1.31 &  5.05 &  4.57 &  --   &  1.27 &  4.34 &  3.94 \\
 NGC2699 &  14 & 1.41 & 146 &  1.83 &  5.86 &  4.28 &  2.40 &  1.76 &  4.76 &  3.58 \\
 NGC2768 &  71 & 0.39 & 205 &  1.70 &  5.02 &  4.08 &  --   &  1.64 &  4.43 &  3.64 \\
 NGC2974 &  24 & 1.04 & 242 &  1.64 &  5.27 &  4.40 &  2.25 &  1.71 &  5.14 &  4.09 \\
 NGC3032 &  17 & 1.19 & 100 &  4.57 &  4.58 &  1.73 &  1.62 &  3.98 &  4.02 &  2.14 \\
 NGC3156 &  25 & 0.80 &  61 &  3.92 &  3.80 &  1.60 &  1.60 &  3.00 &  3.66 &  1.79 \\
 NGC3377 &  38 & 0.52 & 142 &  1.85 &  4.95 &  3.83 &  2.09 &  1.86 &  4.37 &  3.20 \\
 NGC3379 &  42 & 0.64 & 215 &  1.43 &  5.23 &  4.57 &  2.20 &  1.44 &  4.81 &  4.15 \\
 NGC3384 &  27 & 0.75 & 158 &  1.94 &  5.92 &  4.16 &  2.47 &  1.84 &  5.12 &  3.70 \\
 NGC3414 &  33 & 0.60 & 236 &  1.44 &  5.08 &  4.48 &  2.15 &  1.50 &  4.49 &  3.72 \\
 NGC3489 &  19 & 1.05 & 100 &  2.82 &  4.99 &  2.84 &  2.01 &  2.60 &  4.40 &  2.60 \\
 NGC3608 &  41 & 0.49 & 189 &  1.49 &  5.30 &  4.36 &  --   &  1.60 &  4.73 &  3.75 \\
 NGC4150 &  15 & 1.39 &  78 &  3.47 &  4.23 &  2.17 &  1.69 &  2.87 &  4.11 &  2.35 \\
 NGC4262 &  10 & 2.06 & 199 &  1.42 &  5.07 &  4.54 &  2.09 &  1.49 &  4.46 &  3.90 \\
 NGC4270 &  18 & 1.09 & 137 &  1.77 &  5.09 &  3.45 &  --   &  1.74 &  4.67 &  3.24 \\
 NGC4278 &  32 & 0.74 & 248 &  1.23 &  4.78 &  4.67 &  2.04 &  1.41 &  4.59 &  4.21 \\
 NGC4374 &  71 & 0.43 & 294 &  1.40 &  5.19 &  4.52 &  2.22 &  1.42 &  4.66 &  4.08 \\
 NGC4382 &  67 & 0.38 & 179 &  2.16 &  5.14 &  3.45 &  2.12 &  1.98 &  4.72 &  3.36 \\
 NGC4387 &  17 & 1.16 &  96 &  1.60 &  5.03 &  3.92 &  2.18 &  1.51 &  4.42 &  3.67 \\
 NGC4458 &  27 & 0.74 &  95 &  1.63 &  4.49 &  3.73 &  1.90 &  1.59 &  3.97 &  3.27 \\
 NGC4459 &  38 & 0.71 & 180 &  2.16 &  5.39 &  3.91 &  2.27 &  1.85 &  4.68 &  3.43 \\
 NGC4473 &  27 & 0.92 & 198 &  1.49 &  5.55 &  4.59 &  2.33 &  1.50 &  4.88 &  4.03 \\
 NGC4477 &  47 & 0.43 & 168 &  1.65 &  5.15 &  4.17 &  2.21 &  1.59 &  4.71 &  3.84 \\
 NGC4486 & 105 & 0.29 & 317 &  1.14 &  5.30 &  5.12 &  2.22 &  1.15 &  4.70 &  4.62 \\
 NGC4526 &  40 & 0.66 & 240 &  1.86 &  5.59 &  4.32 &  2.29 &  1.62 &  4.93 &  4.16 \\
 NGC4546 &  22 & 0.94 & 223 &  1.54 &  5.49 &  4.59 &  2.27 &  1.54 &  4.64 &  3.92 \\
 NGC4550 &  14 & 1.28 &  81 &  2.14 &  4.57 &  3.15 &  2.00 &  1.99 &  4.32 &  2.93 \\
 NGC4552 &  32 & 0.63 & 277 &  1.39 &  5.93 &  5.04 &  2.31 &  1.37 &  5.33 &  4.55 \\
 NGC4564 &  21 & 1.02 & 173 &  1.52 &  5.98 &  4.79 &  2.42 &  1.55 &  4.85 &  4.00 \\
 NGC4570 &  14 & 1.43 & 199 &  1.45 &  5.87 &  4.70 &  2.34 &  1.45 &  4.79 &  3.99 \\
 NGC4621 &  46 & 0.56 & 229 &  1.40 &  5.53 &  4.73 &  2.28 &  1.43 &  4.75 &  4.14 \\
 NGC4660 &  11 & 1.83 & 229 &  1.43 &  5.87 &  4.88 &  2.34 &  1.47 &  4.80 &  4.07 \\
 NGC5198 &  25 & 0.80 & 209 &  1.49 &  5.50 &  4.66 &  --   &  1.56 &  4.70 &  3.93 \\
 NGC5308 &  10 & 2.04 & 252 &  1.46 &  5.42 &  4.57 &  --   &  1.46 &  4.92 &  4.23 \\
 NGC5813 &  52 & 0.39 & 231 &  1.51 &  5.16 &  4.58 &  2.15 &  1.52 &  4.74 &  4.12 \\
 NGC5831 &  35 & 0.67 & 166 &  1.77 &  5.55 &  4.09 &  2.35 &  1.78 &  4.78 &  3.32 \\
 NGC5838 &  23 & 0.87 & 290 &  1.64 &  5.86 &  4.55 &  2.35 &  1.58 &  5.05 &  4.19 \\
 NGC5845 &   5 & 4.45 & 285 &  1.51 &  6.12 &  4.80 &  2.38 &  1.52 &  5.38 &  4.41 \\
 NGC5846 &  81 & 0.29 & 237 &  1.34 &  5.50 &  4.85 &  --   &  1.32 &  4.89 &  4.38 \\
 NGC5982 &  27 & 0.94 & 269 &  1.63 &  5.96 &  4.50 &  --   &  1.60 &  5.18 &  3.98 \\
 NGC7332 &  11 & 1.91 & 135 &  2.28 &  5.93 &  3.71 &  2.30 &  2.16 &  4.80 &  3.17 \\
 NGC7457 &  65 & 0.39 &  62 &  2.28 &  4.81 &  2.88 &  2.02 &  2.25 &  4.33 &  2.68 \\
\hline
 \end{tabular}

\medskip 

{\em Notes:}\/ (1) NGC number. (2) Effective (half-light) radius
$R_{\rmn e}$ measured from HST/WFPC2 + MDM images. (3) Ratio between
the maximum radius $R_{\rmn max}$ sampled by the \sauron\/ observations
and $R_{\rmn e}$. The maximum radius is calculated as the equivalent
circular radius matching the area of the full \sauron\/ field of view.
Note, that for five galaxies the values given here are slightly smaller
than the ones given in Paper~IV since we have removed bad bins for the
line-strength indices (see also Section~\ref{sec:maps}). (4) Velocity
dispersion of the luminosity weighted spectrum within a circular
aperture of \reb.  (5)~--~(8) Fully corrected line strength index
measurements of the luminosity weighted spectrum within \reb\/ for the
\hb, Fe5015, \mgb\/ and \fes\/ indices. Due to limited field coverage
we cannot determine the \fes\/ indices for eleven galaxies. (9)~-~(11)
Fully corrected line strength index measurements of the luminosity
weighted spectrum within $R_{\rmn e}$ for the \hb, Fe5015, and \mgb\/
indices. For galaxies with less than one $R_{\rmn e}$ coverage we
applied the aperture corrections given in Table~\ref{tab:aper}. Formal
errors of the line strength indices are below 0.1\,\AA. We note,
however, that there are systematic errors which we estimate to be of
the order of 0.06, 0.15, 0.08, 0.06\,\AA\/ for the \hb, Fe5015, \mgb,
and \fes\/ indices, respectively. For the velocity dispersion we adopt
an error of 5\%.
\end{minipage}


\end{center}
\end{table*}

\begin{figure}
 \includegraphics[width=84mm]{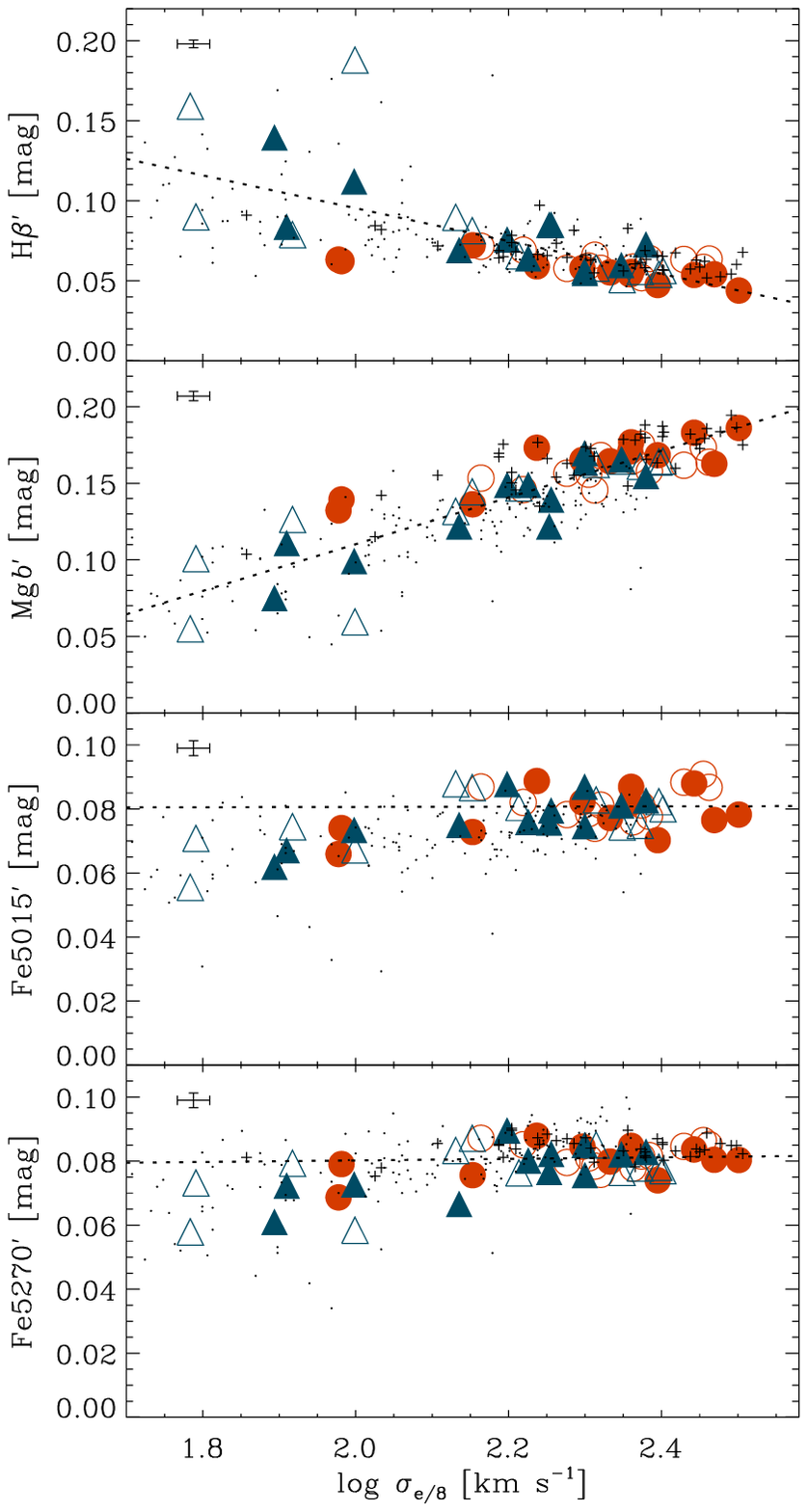}
 \caption{\re\/ aperture line strength index {\it
     versus}\/ $\sigma$ relations. All indices are given in magnitudes.
   The \fes\/ index was converted to the Fe5270 index by using
   Equation~\ref{eq:fe5270s}. The filled and open circles are cluster
   and field ellipticals, respectively; filled and open triangles are
   cluster and field S0s, respectively. Mean error bars reflecting
   mostly systematic errors are shown in the upper left corner of each
   panel. Small dots are taken from \citet{cald03}. The small plus
   signs represent the data of \citet{gon93}. Data from the literature
   with $\sigma < 50$\,\kms\/ are not shown. The dashed lines are fits
   to the data for all galaxies with $\log \sigma \ge 2.1$. The results
   of the linear fits are summarized in Table~\ref{tab:fit}.}
 \label{fig:indexsigma}
\end{figure}

Our main result for the \re\/ extractions is shown in
Figure~\ref{fig:indexsigma}. For the index--$\sigma$ relations all
indices are expressed in magnitudes (see Section~\ref{sec:ls_grad}).
The \mgb\/ index shows a tight correlation with $\log \sigma$ which has
been observed by many authors
\citep[e.g.,][]{ter81,bur88,ben93,coll99,ber03,wor03,den05}. The
H$\beta$ index shows a relatively tight, negative correlation for
galaxies with velocity dispersions $\log \sigma \ge 2.1$. For lower
velocity dispersions there is evidence for an increased scatter
\citep[see also e.g.,][]{kun00,cald03}. The deviations from the main
relation for galaxies with $\log \sigma < 2.1$ clearly anti-correlate
in the \mgb, \hb--$\sigma$ relations.

The two Fe indices show roughly constant index values for velocity
dispersions $\log \sigma \ge 2.1$. Similar to the \mgb\/ and
\hb--$\sigma$ relations we also find an anti-correlation with respect
to the fitted relation between Fe and \hb\/ index strength at a given
$\sigma$ albeit at lower significance. At lower velocity dispersions we
find there is weak evidence for an increasing index strength with
increasing velocity dispersion. Similar trends have been noticed by
\citet{proc02} in an analysis of spiral bulges and E/S0s. The absence
of a clear correlation between Fe indices and the central velocity
dispersion has been noted by several authors
\citep[e.g.,][]{fis96,jor99} and we can confirm this for our sample.
However, for other samples a significant correlation between Fe indices
and the central velocity dispersion has been found
\citep[e.g.,][]{kun00,cald03,ber03}.

In order to compare our data with the literature we choose two datasets
which cover luminous early-type galaxies \citep[][\re\/
extractions]{gon93,tra00a} and fainter ones \cite[][global
extractions]{cald03}. Our sample of 48 early-type galaxies shows good
agreement with the literature for the \hb, \mgb\/ and Fe5270 indices
(see Figure~\ref{fig:indexsigma}). For Fe5015, which was only observed
by \citet{cald03}, an offset of approximately 0.01\,mag at intermediate
velocity dispersions is visible, in the sense that our data has the
larger values. The best fitting linear index--$\sigma$ relations for
$\log \sigma \ge 2.1$, using errors in both variables are given in
Table~\ref{tab:fit}.

\begin{table}
  \centering
  \begin{minipage}{71mm}
    \caption{Linear fits to index--$\sigma$ relations (for $\log \sigma \ge 2.1$)}
    \label{tab:fit}
    \begin{tabular}{ll}
      \hline 
\hb$^\prime$    & $= -0.103(\pm0.017) \log \sigma + 0.301(\pm0.039)$ \\
\mgb$^\prime$   & $= +0.153(\pm0.021) \log \sigma - 0.195(\pm0.050)$ \\
Fe5015$^\prime$ & $= +0.000(\pm0.010) \log \sigma + 0.080(\pm0.024)$ \\
Fe5270$^\prime$ & $= +0.002(\pm0.010) \log \sigma + 0.075(\pm0.024)$ \\
\hline
\end{tabular}

\medskip

  Notes: The linear fitting was performed by taking into account errors
  in both variables. The error in the intercepts and slopes have been
  derived by scaling the errors in the observables by a constant factor
  until the $\chi^2$ probability is $\ge 0.5$. 
\end{minipage}
\end{table}

There is no significant evidence for an environmental influence on the
index--$\sigma$ relations in our sample. However, it is generally the
low velocity dispersion S0s which show the strongest anti-correlation
between \hb\/ and metal line strengths. The reason for the scatter in
the index--$\sigma$ relations is typically ascribed to a combination of
age, metallicity and abundance ratio variations at a given $\sigma$
\citep[e.g.,][]{kun01,tho05}. We explore this further in a forthcoming
paper in this series.


%
%
\section{Concluding remarks}
\label{sec:conclusions}
The maps presented in this paper are the result of a comprehensive
survey of the absorption line strength distributions of nearby
early-type galaxies with an integral-field spectrograph. This data set
demonstrates that many nearby early-type galaxies display a significant
and varied structure in their line strength properties. This structure
is sometimes very apparent as in the case of the post-starburst
affected central regions of galaxies, or it can be of subtle nature as
seen in the deviations of \mgb\/ isoindex contours compared to the
isophotes of galaxies.

The two-dimensional coverage of the line strengths allows us to connect
the stellar populations with the kinematical structure of the galaxies
and thus improve our knowledge of the star-formation and assembly
history of early-type galaxies. For example, we find the \mgb\/
isoindex contours to be flatter than the isophotes for 18 galaxies
which generally exhibit significant rotation. We infer from this that
the rotational supported sub-structure features a higher metallicity
and/or an increased Mg/Fe ratio as compared to the galaxy as a whole.
Further steps in this direction will be presented in forthcoming papers
of this series.

The metal line strength maps show generally negative gradients with
increasing radius, while the \hb\/ maps are typically flat or show a
mild positive outwards radial gradient. A few galaxies show strong
central peaks and/or elevated overall \hb-strength likely connected to
recent star-formation activity. For the most prominent post-starburst
galaxies even the metal line strength maps show a reversed gradient.
We use the maps to compute average line strengths integrated over
circular apertures of one-eighth effective radius. The resulting index
versus velocity dispersion relations compare well with previous
long-slit work.

%
%
\renewcommand{\thefigure}{\arabic{figure}\alph{subfigure}}
\setcounter{subfigure}{1}

\begin{figure*}
\begin{center} 
  \includegraphics[draft=false,scale=0.99,trim=0cm 0cm
    0cm 0cm]{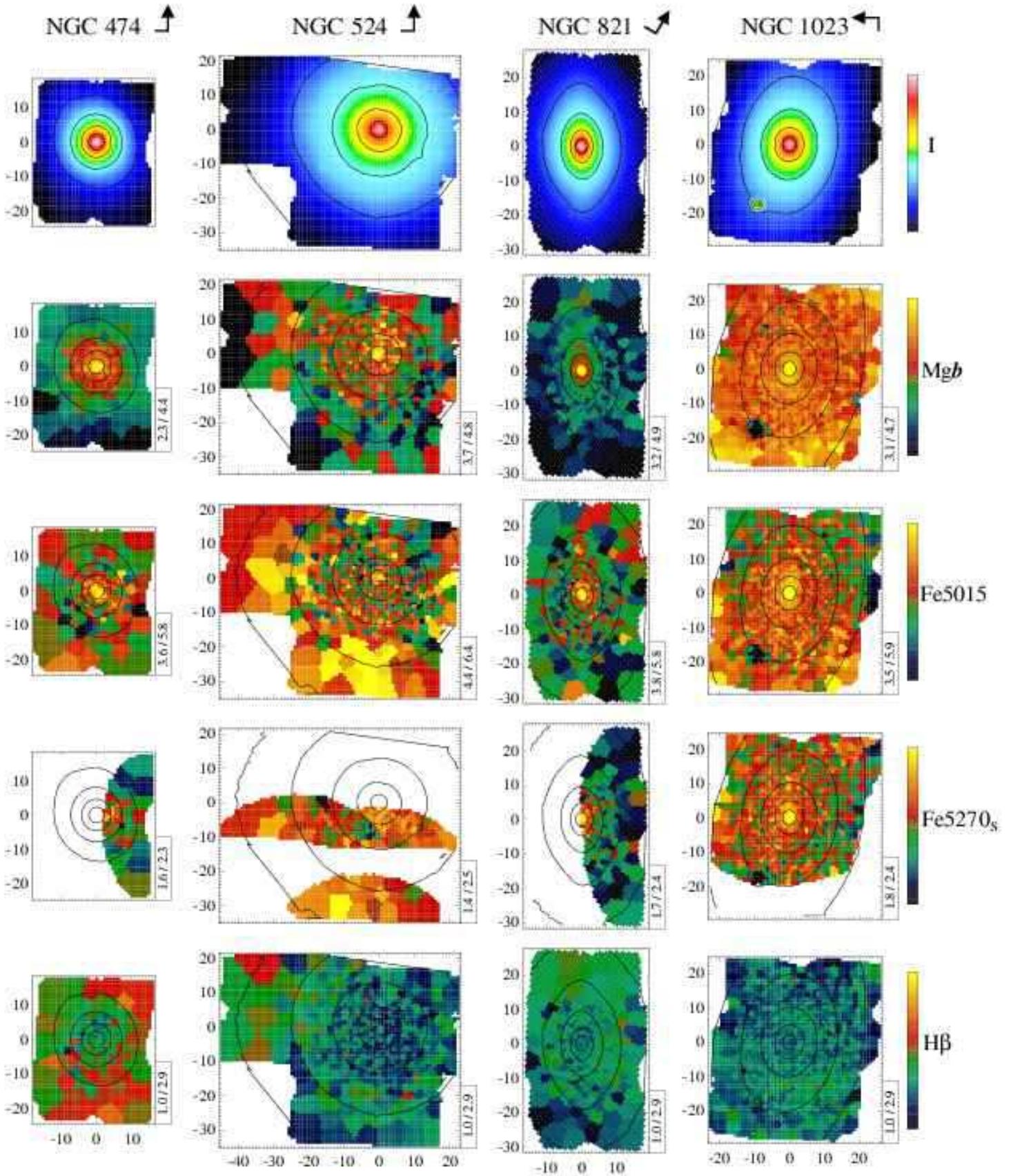} 
\end{center} 
\caption[]{Maps of the line strength the 48 E and S0 galaxies
  in the \sauron\ representative sample. The \sauron\ spectra have been
  spatially binned to a minimum signal-to-noise of 60 by means of the
  centroidal Voronoi tessellation algorithm of \protect\citet{cap03}.
  All maps have the same spatial scale. From top to bottom: i)
  reconstructed total intensity; ii) \mgb\/ line strength; iii) Fe5015
  line strength iv) \fes\/ line strength and v) \hb\/ line strength.}
\label{fig:maps1}
\end{figure*}

\addtocounter{figure}{-1}
\addtocounter{subfigure}{1}

\begin{figure*}
\begin{center} 
  \includegraphics[draft=false,scale=0.99,trim=0cm 0.cm
    0cm 0cm]{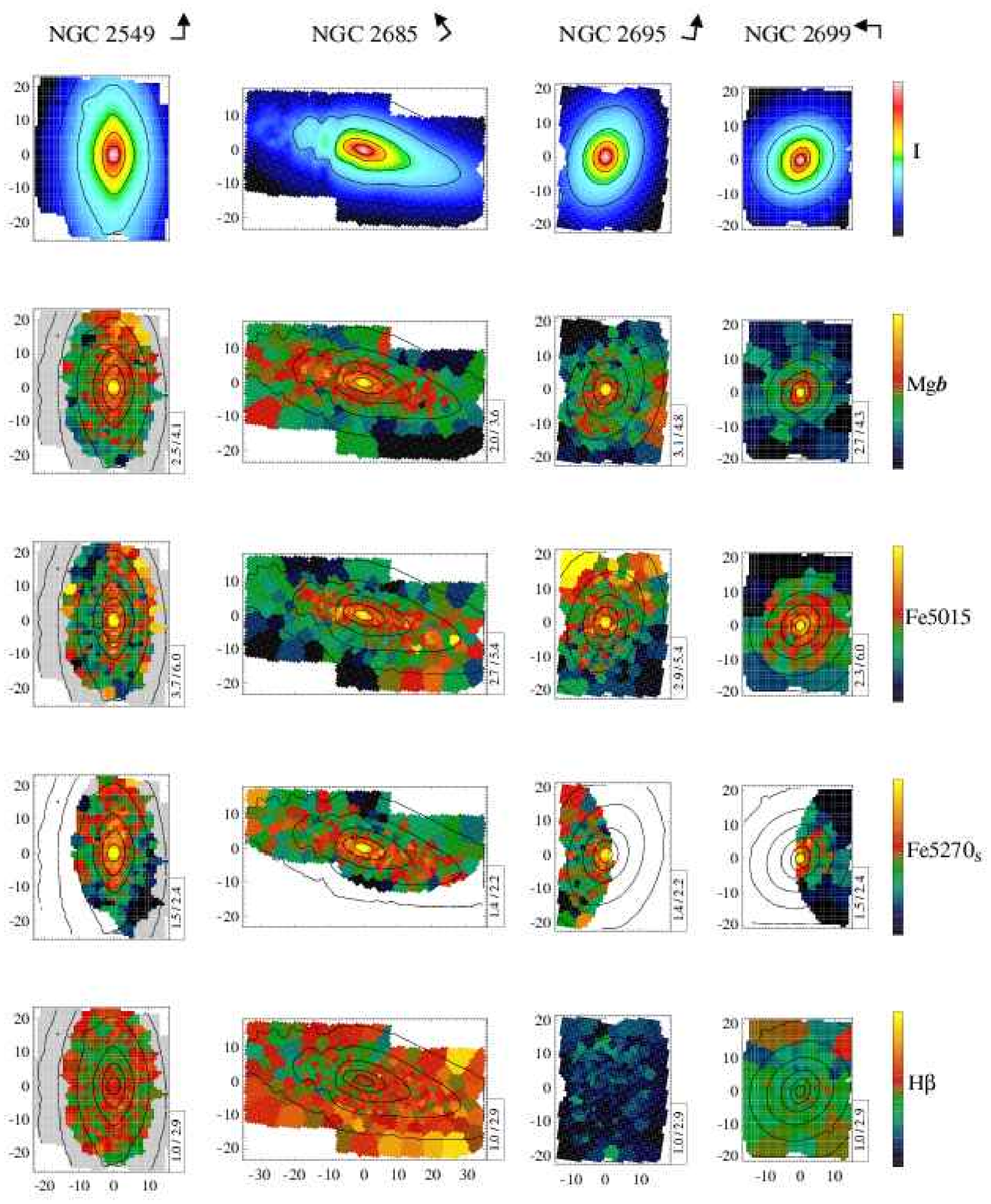} 
\end{center} 
\caption[]{}
\label{fig:maps2}
\end{figure*}

\addtocounter{figure}{-1}
\addtocounter{subfigure}{1}

\begin{figure*}
\begin{center} 
  \includegraphics[draft=false,scale=0.99,trim=0cm 0.cm
    0cm 0cm]{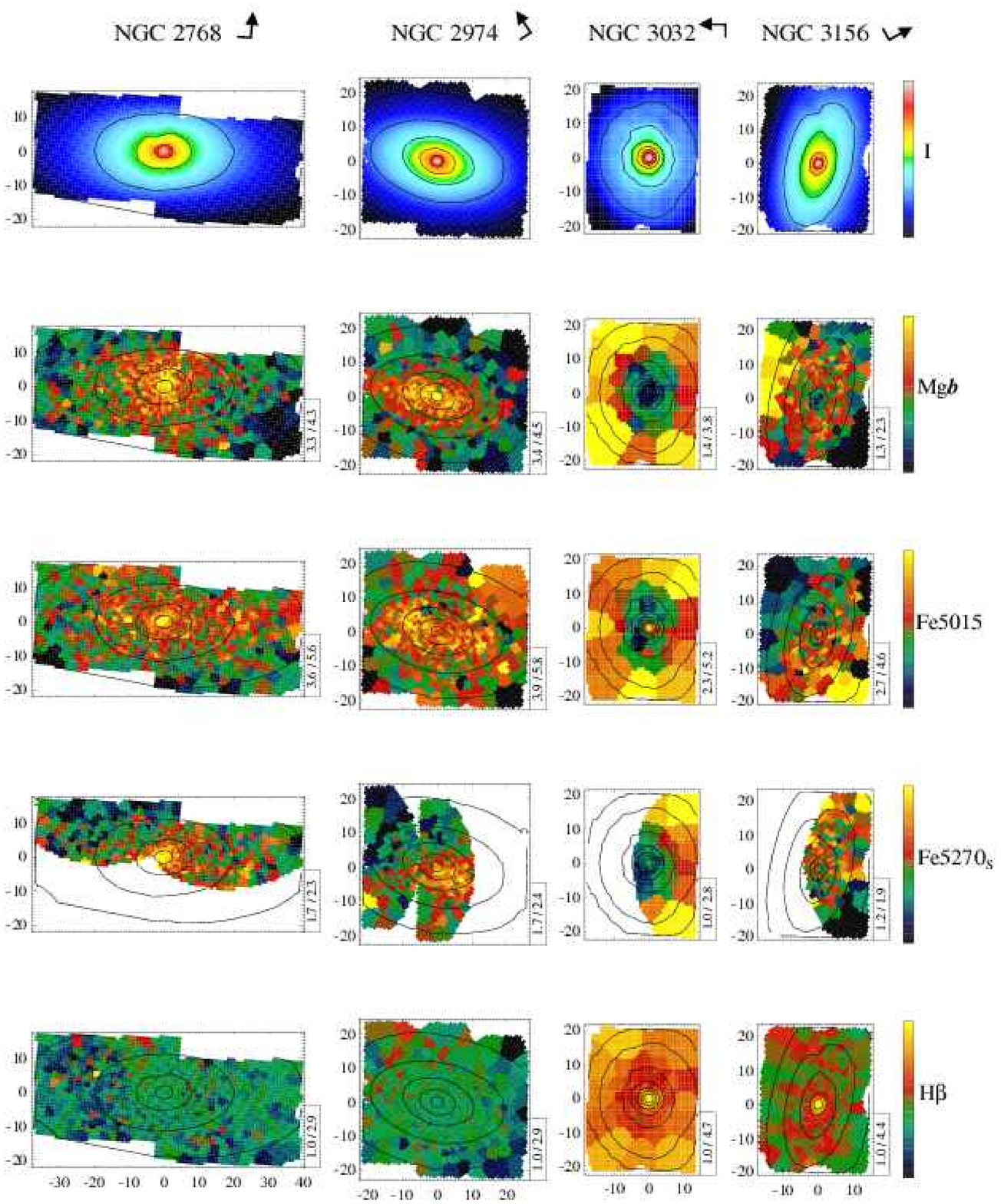} 
\end{center} 
\caption[]{}
\label{fig:maps3}
\end{figure*}

\addtocounter{figure}{-1}
\addtocounter{subfigure}{1}

\begin{figure*}
\begin{center} 
  \includegraphics[draft=false,scale=0.99,trim=0cm 0.cm
    0cm 0cm]{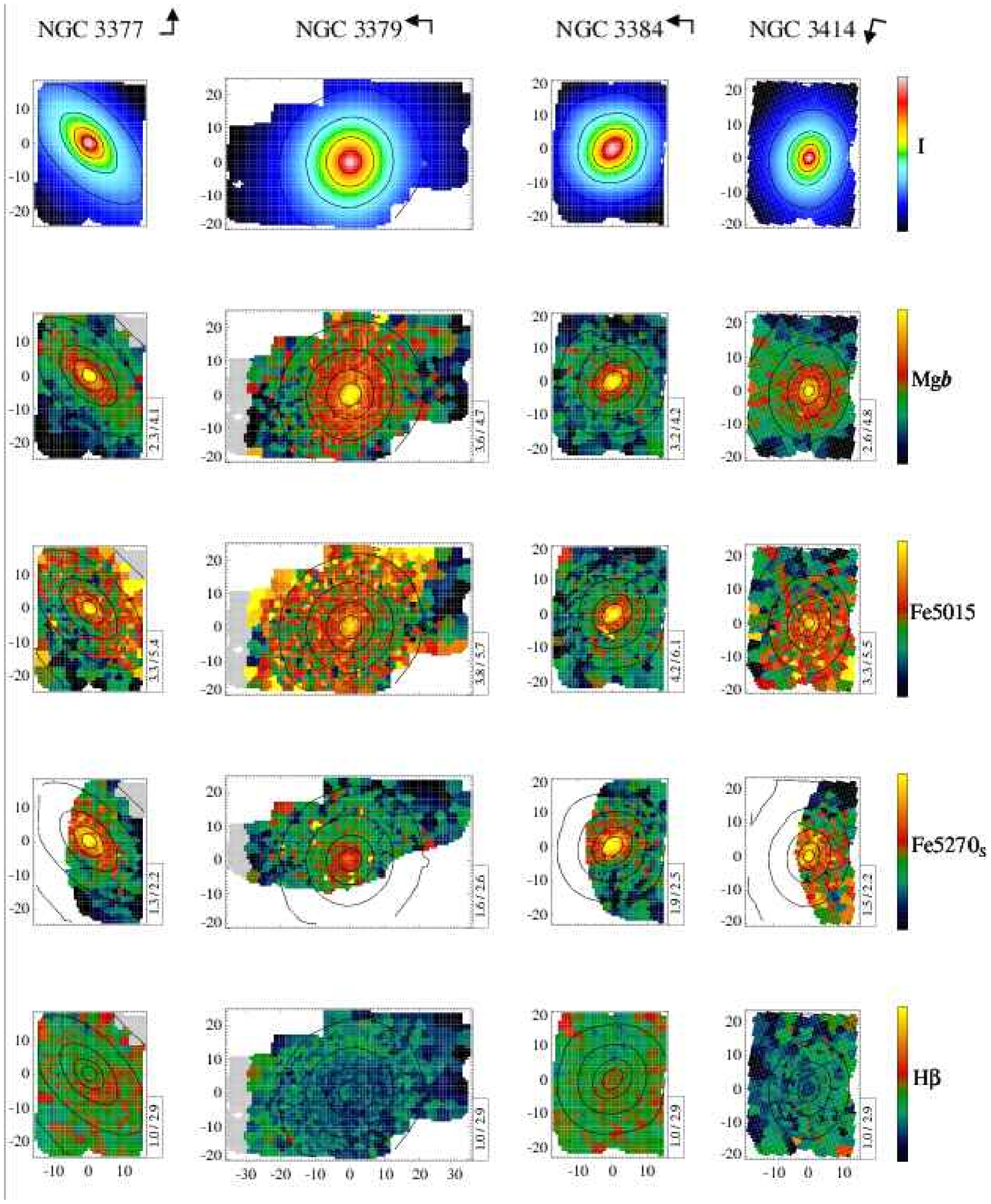} 
\end{center} 
\caption[]{}
\label{fig:maps4}
\end{figure*}

\addtocounter{figure}{-1}
\addtocounter{subfigure}{1}

\begin{figure*}
\begin{center} 
  \includegraphics[draft=false,scale=0.99,trim=0cm 0.cm
    0cm 0cm]{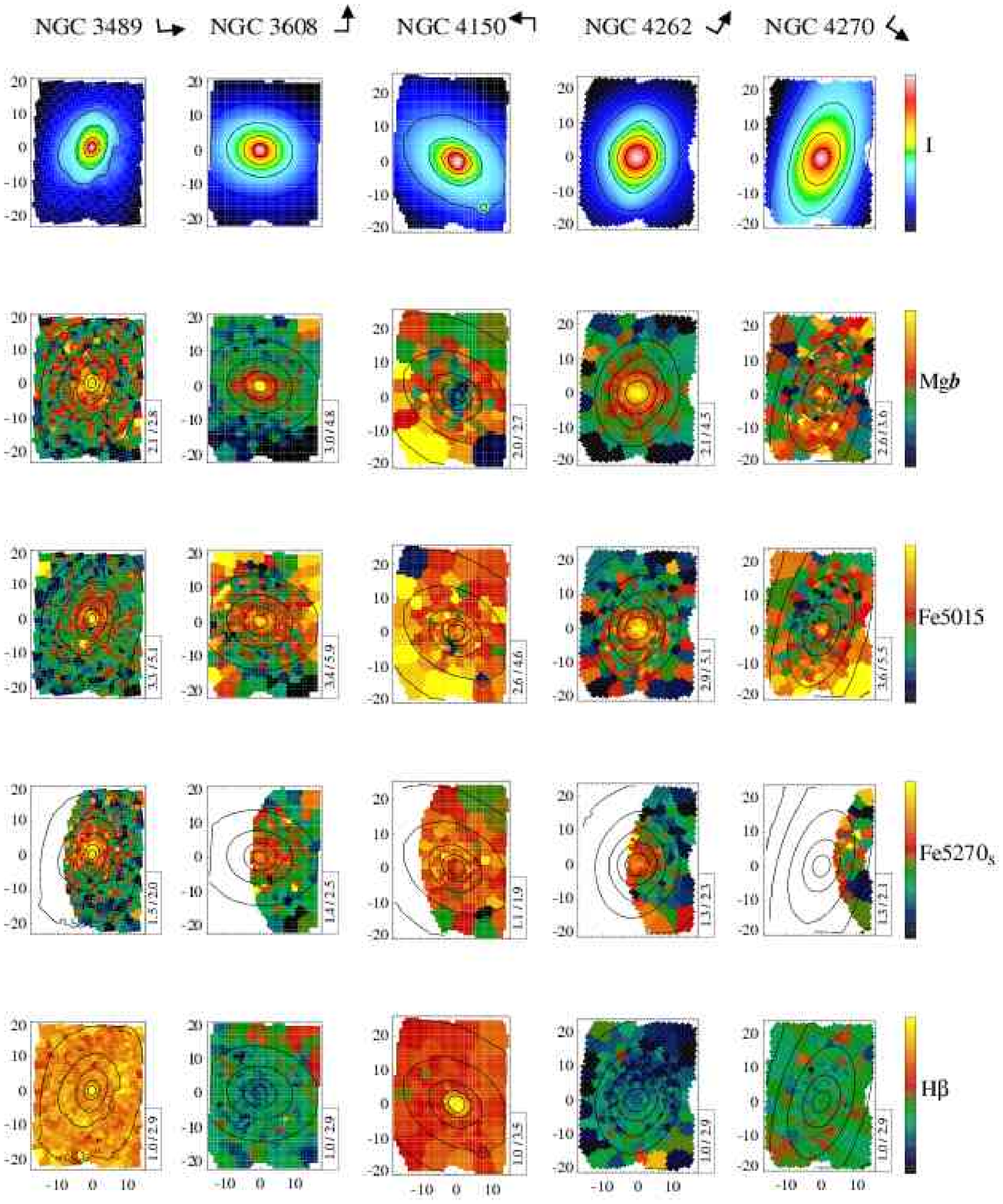} 
\end{center} 
\caption[]{}
\label{fig:maps5}
\end{figure*}

\addtocounter{figure}{-1}
\addtocounter{subfigure}{1}

\begin{figure*}
\begin{center} 
  \includegraphics[draft=false,scale=0.99,trim=0cm 0.cm
    0cm 0cm]{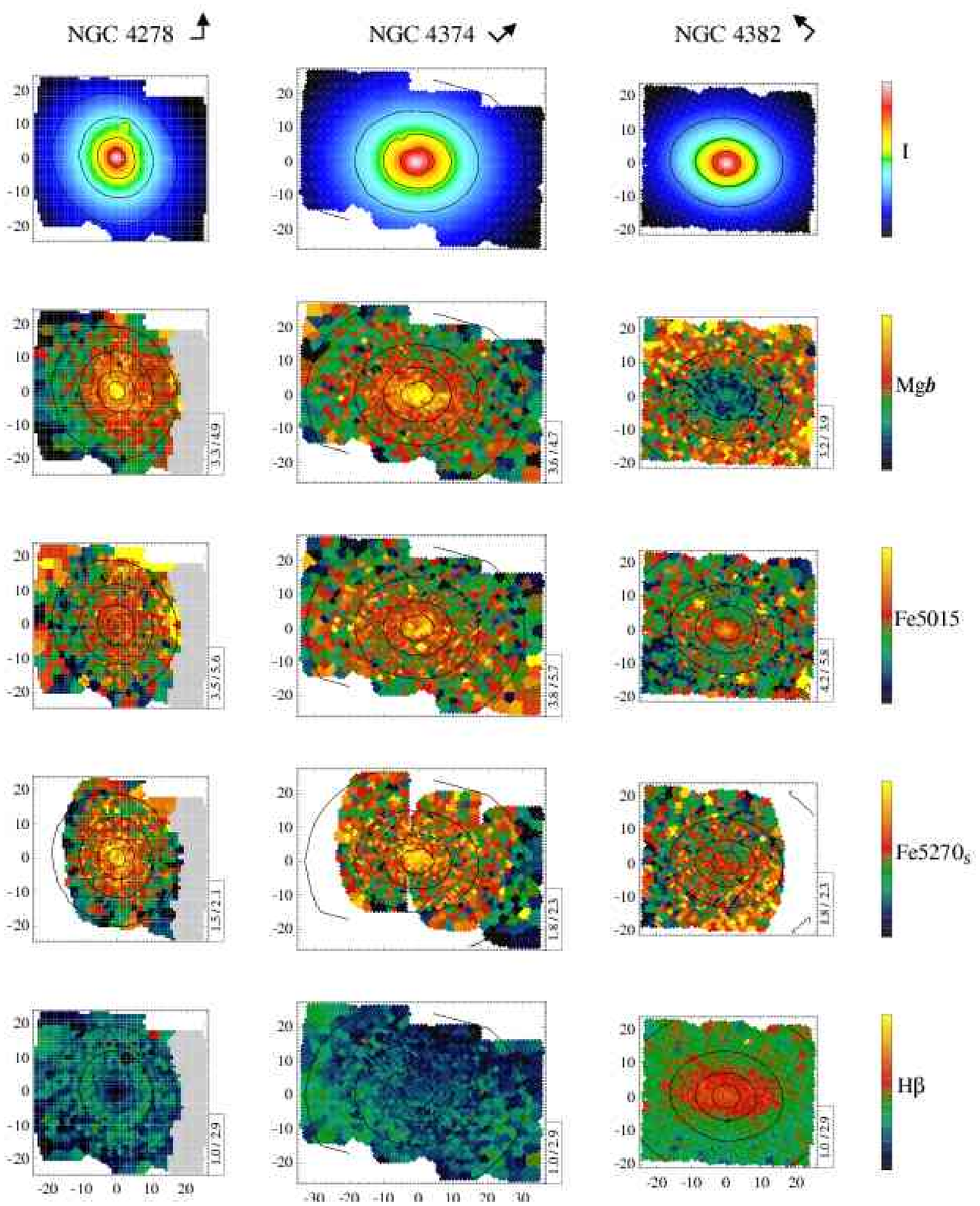} 
\end{center} 
\caption[]{}
\label{fig:maps6}
\end{figure*}

\addtocounter{figure}{-1}
\addtocounter{subfigure}{1}

\begin{figure*}
\begin{center} 
  \includegraphics[draft=false,scale=0.99,trim=0cm 0.cm
    0cm 0cm]{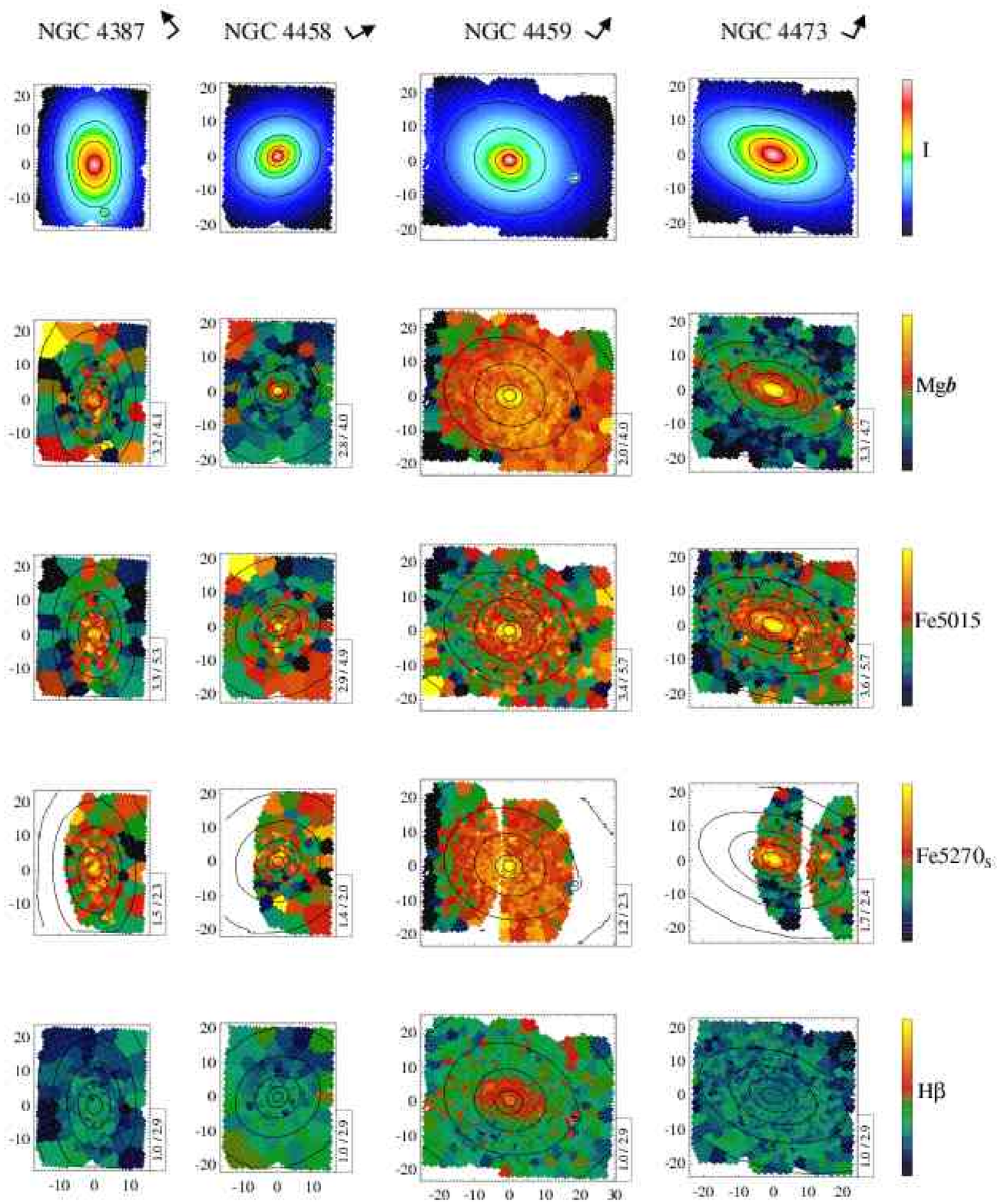} 
\end{center} 
\caption[]{}
\label{fig:maps7}
\end{figure*}

\addtocounter{figure}{-1}
\addtocounter{subfigure}{1}

\begin{figure*}
\begin{center} 
  \includegraphics[draft=false,scale=0.99,trim=0cm 0.cm
    0cm 0cm]{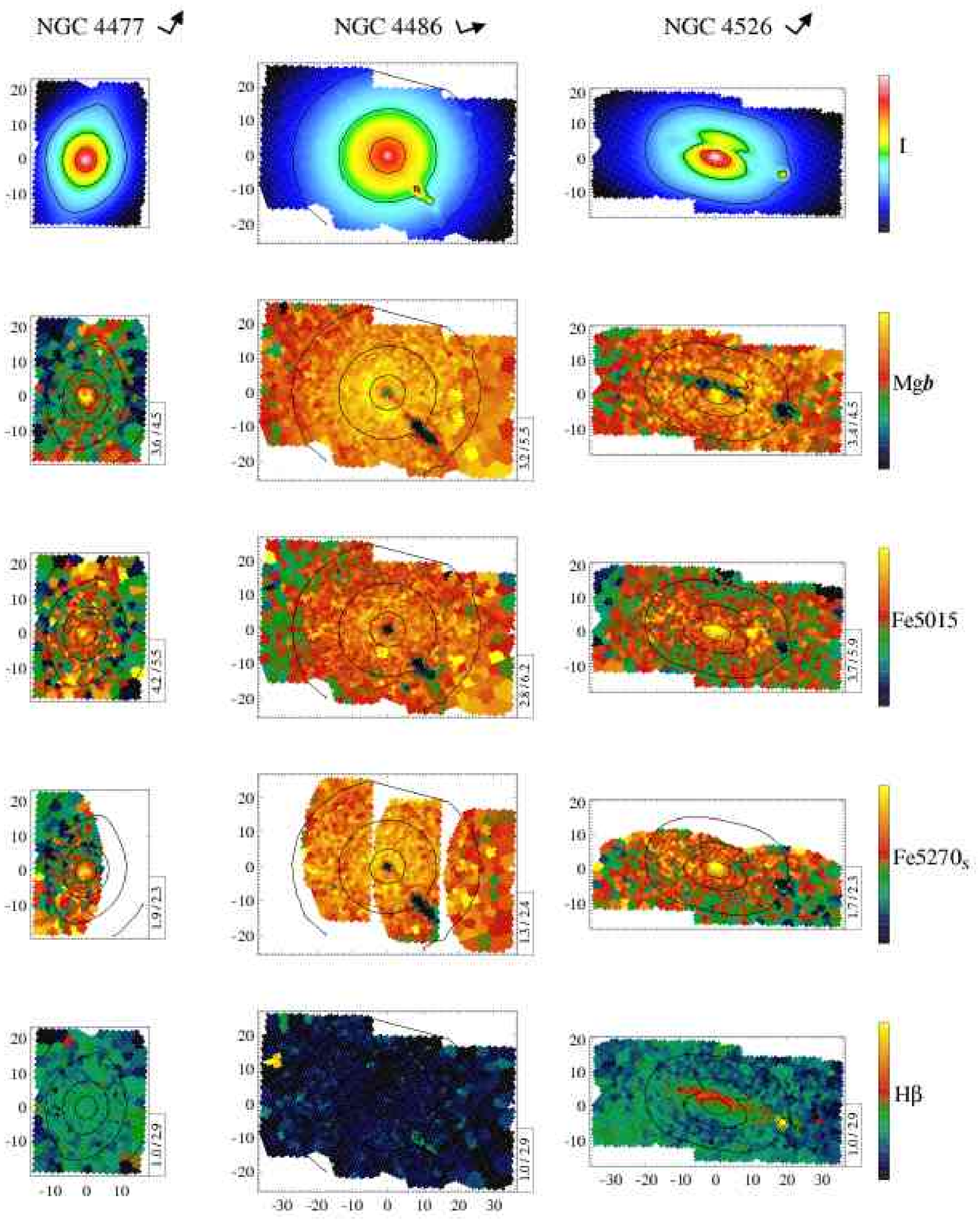} 
\end{center} 
\caption[]{}
\label{fig:maps8}
\end{figure*}

\addtocounter{figure}{-1}
\addtocounter{subfigure}{1}

\begin{figure*}
\begin{center} 
  \includegraphics[draft=false,scale=0.99,trim=0cm 0.cm
    0cm 0cm]{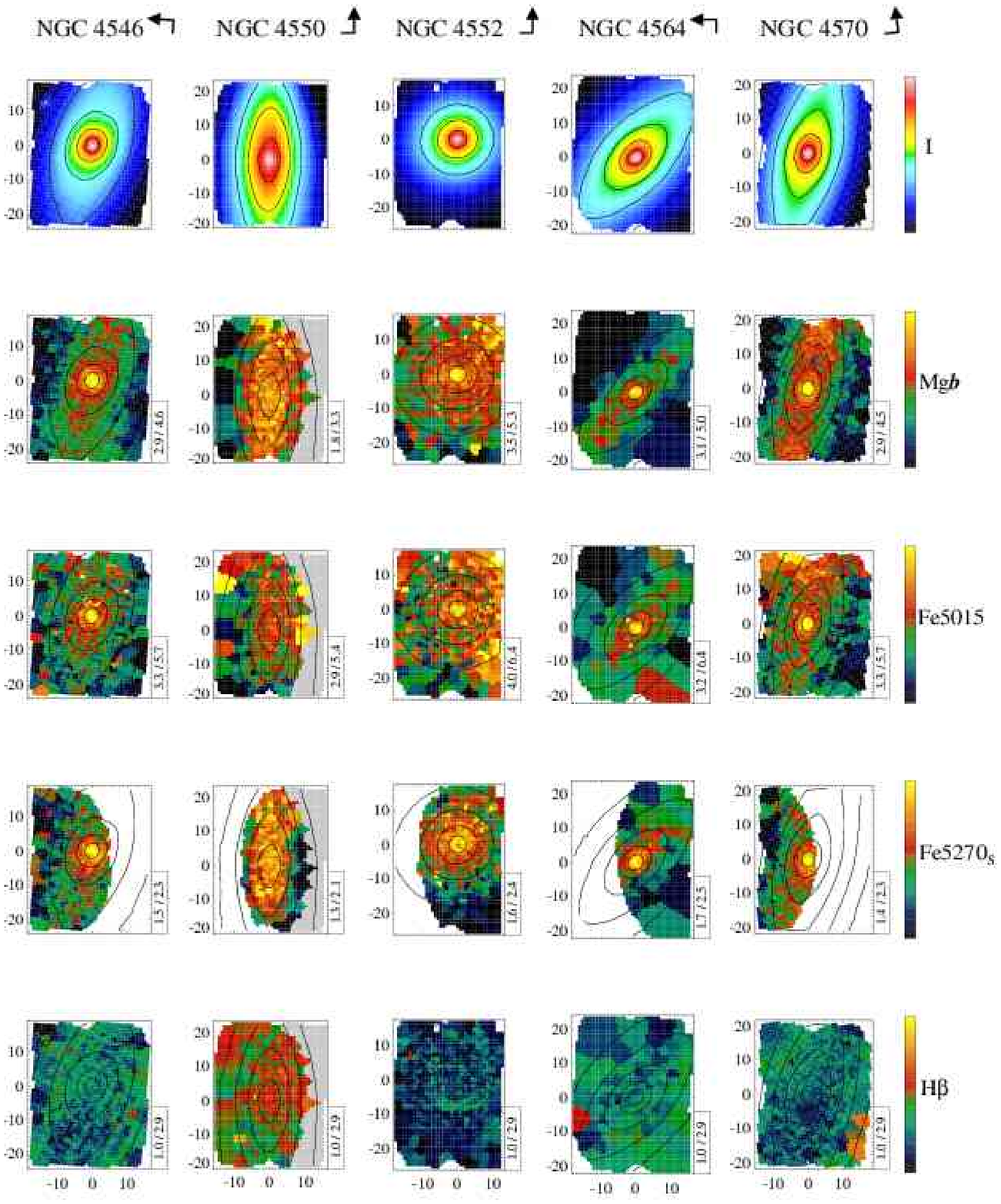} 
\end{center} 
\caption[]{}
\label{fig:maps9}
\end{figure*}

\addtocounter{figure}{-1}
\addtocounter{subfigure}{1}

\begin{figure*}
\begin{center} 
  \includegraphics[draft=false,scale=0.99,trim=0cm 0.cm
    0cm 0cm]{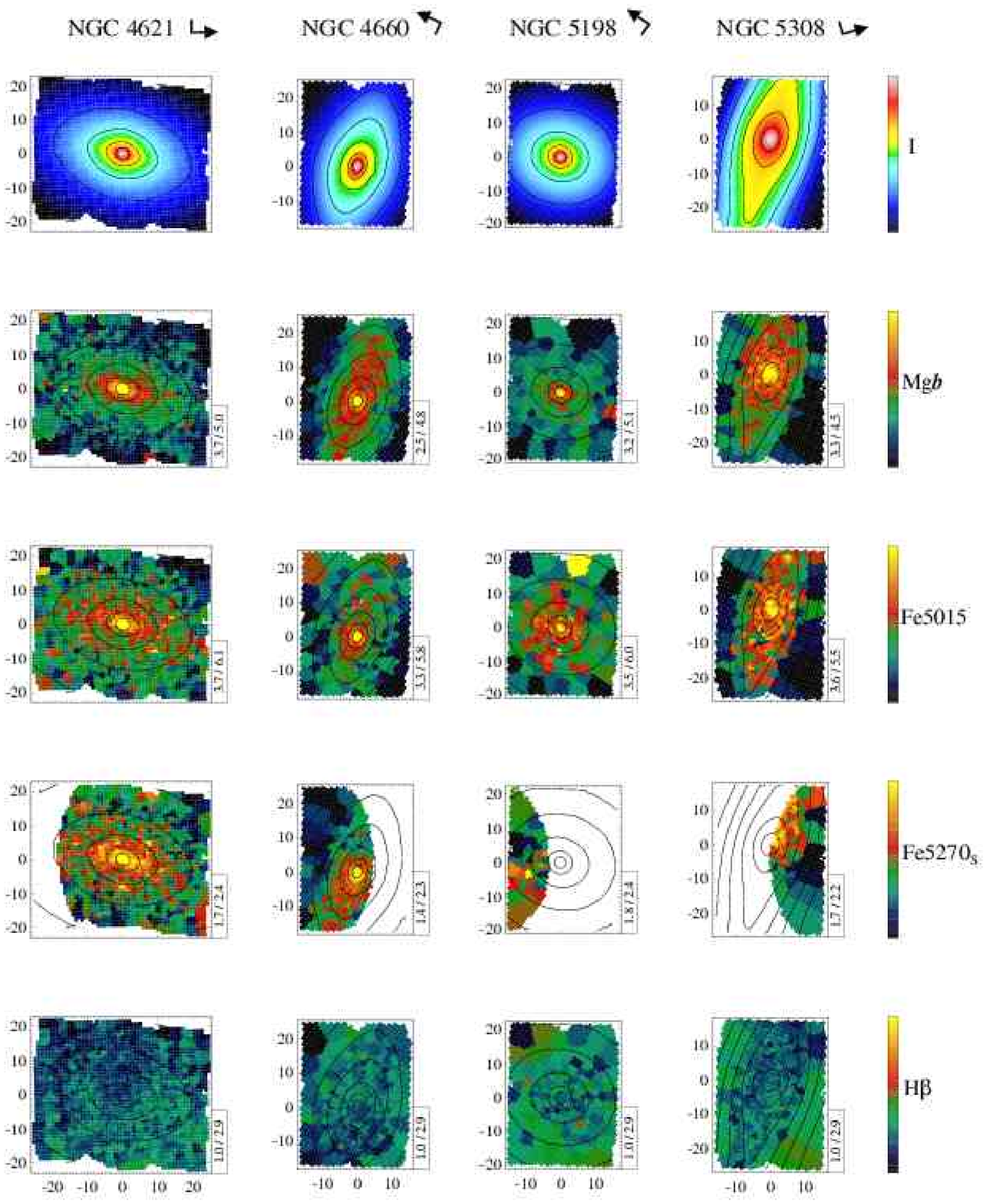} 
\end{center} 
\caption[]{}
\label{fig:maps10}
\end{figure*}

\addtocounter{figure}{-1}
\addtocounter{subfigure}{1}

\begin{figure*}
\begin{center} 
  \includegraphics[draft=false,scale=0.99,trim=0cm 0.cm
    0cm 0cm]{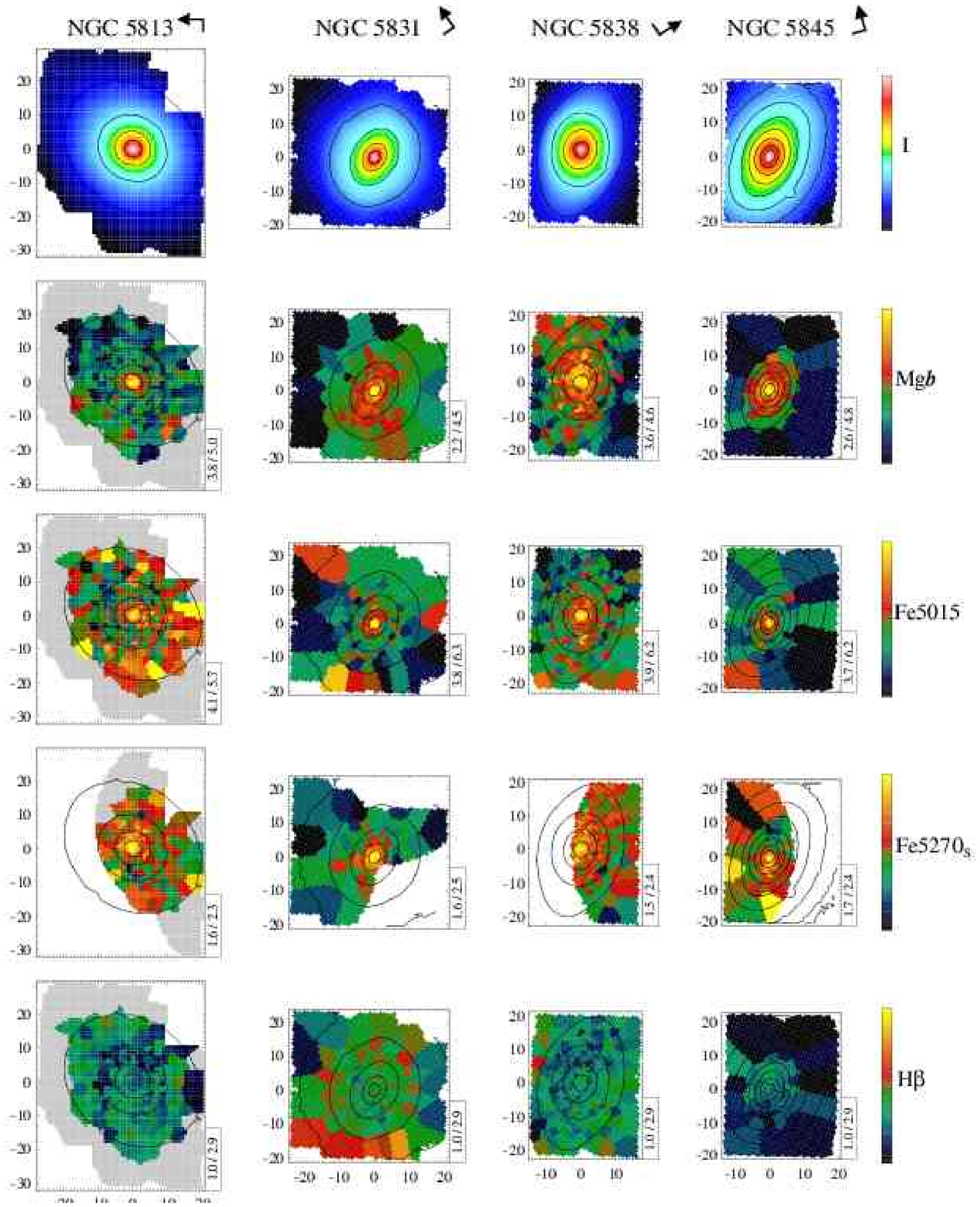} 
\end{center} 
\caption[]{}
\label{fig:maps11}
\end{figure*}

\addtocounter{figure}{-1}
\addtocounter{subfigure}{1}

\begin{figure*}
\begin{center} 
  \includegraphics[draft=false,scale=0.99,trim=0cm 0.cm
    0cm 0cm]{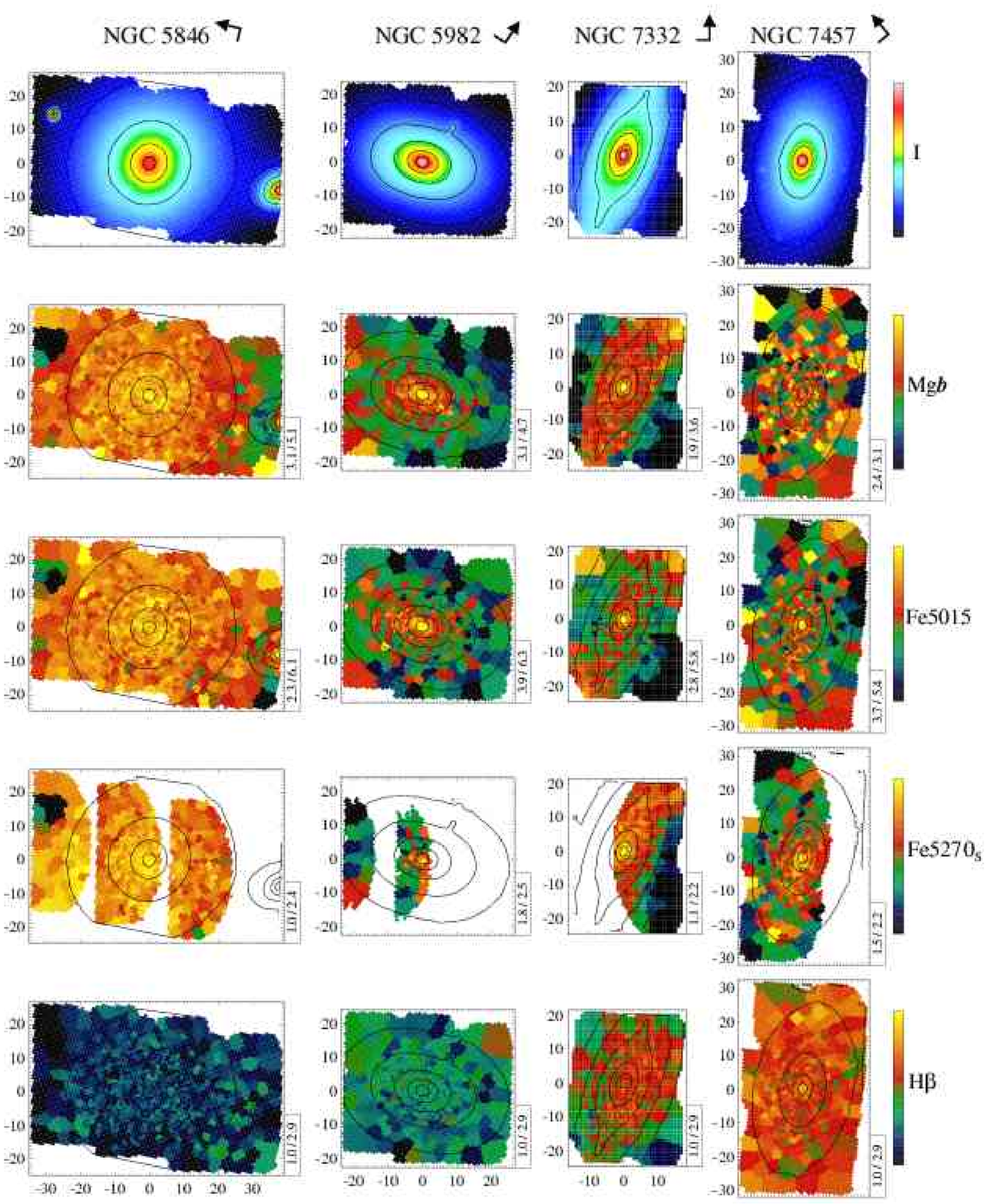} 
\end{center} 
\caption[]{}
\label{fig:maps12}
\end{figure*}

\section*{Acknowledgments}
The \sauron\ project is made possible through grants 614.13.003,
781.74.203, 614.000.301 and 614.031.015 from NWO and financial
contributions from the Institut National des Sciences de l'Univers, the
Universit\'e Claude Bernard Lyon~I, the Universities of Durham, Leiden,
and Oxford, the British Council, PPARC grant `Extragalactic Astronomy
\& Cosmology at Durham 1998--2002', and the Netherlands Research School
for Astronomy NOVA. RLD is grateful for the award of a PPARC Senior
Fellowship (PPA/Y/S/1999/00854) and postdoctoral support through PPARC
grant PPA/G/S/2000/00729. The PPARC Visitors grant (PPA/V/S/2002/00553)
to Oxford also supported this work. We are grateful to the Lorentz
center at university of Leiden for generous hospitality. MC
acknowledges support from a VENI grant 639.041.203 awarded by the
Netherlands Organization for Scientific Research. JFB acknowledges
support from the Euro3D Research Training Network, funded by the EC
under contract HPRN-CT-2002-00305.  This project made use of the
HyperLeda and NED databases. Part of this work is based on data
obtained from the ESO/ST-ECF Science Archive Facility. The \sauron\/
team wishes to thank Emilie Jourdeuil for contributions in an early
stage of this project.

%
%
%

\bibliographystyle{mn2e}
\bibliography{references}

\appendix

\section{Description for individual galaxies}
\label{sec:list}
Here we briefly comment on the line strength structures observed in the
\sauron\ maps of the E/S0 sample presented in this paper. Individual
comments on the stellar kinematics and the emission line maps are
presented in Papers~III and V, respectively. A thorough and
quantitative assessment of the line strength maps with the help of
stellar population models will be carried out in subsequent papers.
Independently derived line strength maps for NGC\,524, NGC\,1023
NGC\,3379, NGC\,3384, NGC\,4550, NGC\,7332 and NGC\,7457 based on our
observations were presented and discussed in \citet{Afa02},
\citet{Sil02}, \citet{Sil03} and \citet{sil05}.

\begin{description} 

\item[\bf NGC\,474:] This galaxy (Arp~227), famous for its shell
  structures \citep[e.g.][]{Tur99}, shows normal metal line strength
  gradients. The \hb\/ map appears relatively flat in the inner regions
  while we observe rising \hb\/ strength towards larger radii.
  
\item[\bf NGC\,524:] This galaxy shows regular \mgb\/ gradients and the
  \hb\/ map shows a mild positive gradient. The two Fe maps show a
  noisy structure which may partly be caused by low S/N in the outer
  bins.
  
\item[\bf NGC\,821:] A close to edge-on galaxy with a rapidly rotating
  disk like component (see Paper~III) shows an enhanced core region in
  all metal line maps. There is also evidence for the \mgb\/ contours
  to be flatter than the isophotes. The \hb\/ map appears rather
  featureless \citep[see also][]{rmd05}. The maps are consistent with
  the results from long-slit spectroscopy of \citet{san04}.
  
\item[\bf NGC\,1023:] This SB0 galaxy, with a prominent twist in the
  central velocity field (see Paper~III), shows a central concentration
  in all metal lines. The metal line strength at larger radii appear
  flat and even tend to rise again at the edges of the field of view.
  The \hb\/ map is consistent with a constant value over the full field
  of view.
  
\item[\bf NGC\,2549:] This galaxy with a thin, rapidly rotating
  component (see Paper~III) shows enhanced Fe5015 and \fes\/ strength
  along this component. The \hb\/ line strength is also elevated in the
  central regions.
  
\item[\bf NGC\,2685:] This famous object, the Helix galaxy
  \citep{Bur59,Pel93}, shows significant extinction due to polar dust
  lanes on its North East side. The \hb\/ map shows enhanced values
  over large regions of the FoV, while the metal line strength maps are
  elevated along the rotation direction.
  
\item[\bf NGC\,2695:] The metal line strength maps show normal gradients
  consistent with the isophotes. The \hb\/ map is flat, however, at
  notably low values. 
  
\item[\bf NGC\,2699:] This galaxy shows regular metal line strength
  gradients consistent with the isophotes. The \hb\/ map is relatively
  flat at a slightly elevated level compared to the bulk of the objects
  in our sample.
  
\item[\bf NGC\,2768:] This galaxy with a rather cylindrical velocity
  field shows regular metal line strength gradients. The \hb\/ map
  appears rather flat. Significant dust extinction is present north of
  the center \citep{mich99}.
  
\item[\bf NGC\,2974:] This rapidly rotating galaxy shows evidence of
  enhanced metal line strength along the rotation direction. The \hb\/
  map appears relatively flat. The long-slit observations of
  \citet{ram05} show declining \hb\/ strength with radius while
  \citet{car93} data suggest a modest increase with radius. Our
  emission line maps (Paper~V) show significant \oiii\/ and \hb\/
  emission basically over the full FoV and thus different \hb\/
  emission corrections are likely the cause of the literature
  disagreement on \hb\/ absorption strengths.
    
\item[\bf NGC\,3032:] This dusty galaxy shows negative \mgb\/ and \fes\/
  line strength gradients. The Fe5015 map shows a ring-like structure
  of weak absorption while the center exhibits strong Fe5015
  absorption. The \hb\/ map reveals a prominent peak in the center and
  shows evidence of rising values at larger radii. The unusual
  structure of the line strength maps and the relatively strong
  emission detected in this galaxy (see Figure~\ref{fig:o3hb} and
  Paper~V) suggest that a recent starburst has occurred in this galaxy
  with weak evidence of ongoing star formation.
  
\item[\bf NGC\,3156:] Another dusty galaxy with negative \mgb\/
  line strength gradients in the center. The Fe5015 and \fes\/ maps do
  not show a clear structure. The \hb\/ map exhibits a strong peak in
  the center with a gradual decline to the outskirts. The line strength
  maps suggest a post-starburst scenario for this galaxy \citep[see
  also][]{den05}.
    
\item[\bf NGC\,3377:] This disky galaxy \citep{ben88,mich88} shows metal
  line strengths which are more flattened than the isophotes.
  Conversely, the \hb\/ absorption has a rather constant moderate value
  across the field. The line strength maps are consistent with
  \citet{san04}.
  
\item[\bf NGC\,3379 (M105):] This rather round object shows normal metal
  line strengths gradients consistent with the isophotes. The \hb\/ map
  shows weakly increasing line strength towards larger radii, as found
  by other authors \cite[e.g.,][see also
  Figure~\ref{fig:n3379_comp}]{dav93,san04}.

\item[\bf NGC\,3384:] This SB0 galaxy shows a dynamically cold component
  in the central 5\arcsec\/ (see Paper~III), which may correspond to
  the red major axis disk component found by \citet[][]{bus96}. The
  line strength indices were already presented in Paper~II. The metal
  line strengths show regular gradients consistent with the isophotes.
  The \hb\/ absorption map, at moderately elevated values \citep[see
  also][]{kun01}, appears flat in the inner regions with weak evidence
  for increasing \hb\/ strength at large radii.
  
\item[\bf NGC\,3414:] This disturbed galaxy (Arp~162) contains a
  kinematically decoupled component (see Paper~III) within the central
  10\arcsec. The line strength maps appear to have regular gradients.
  
\item[\bf NGC\,3489:] This dusty galaxy shows a fast-rotating central
  component and a complex morphology of the odd higher-order kinematic
  moment $h_3$ (see Paper~III). The \hb\/ absorption strength is
  globally high \citep[see also][]{kun01,ram05}, with a sharp central
  (2\arcsec) enhancement, also evident with OASIS \citep[][]{mcd04}.
  
\item[\bf NGC\,3608:] This galaxy shows a kinematically decoupled core
  (Paper~III and references therein) extending to around 13\arcsec. Our
  metal line strengths appear more flattened than the isophotes within
  this region. The \hb\/ absorption map shows a mild positive gradient
  with radius. Line strength are consistent with \citet{san04}.
 
\item[\bf NGC\,4150:] Noticeable obscuration from dust is visible in the
  central parts of this galaxy \citep[][]{qui00}. A kinematically
  decoupled component resides in the central few arcseconds of this
  galaxy (see Paper~III), coincident with a strong drop in the \mgb\/
  absorption strength and a corresponding peak in \hb\/ absorption
  strength, which itself is globally enhanced. The iron line maps are
  relatively featureless by comparison, with no strong central
  features. The core is therefore dominated by a young stellar
  population, as found by other authors \citep[e.g.,][]{fer04}.
  
\item[\bf NGC\,4262:] This strongly barred object shows inconspicuous
  metal line gradients. The \hb\/ absorption map shows a mild increase
  with radius.
  
\item[\bf NGC\,4270:] The central peaked metal absorption line strengths
  in this galaxy show evidence of an increase beyond 10\arcsec\/ along
  the major axis, giving a small dip around 4-5\arcsec\/ from the
  center. The \hb\/ absorption map shows no significant features.
  
\item[\bf NGC\,4278:] This LINER \citep{ho97} and radio source
  \citep[e.g.,][]{cap00} shows some central dust features
  \citep[e.g.,][]{pen02}. There is some evidence that the \mgb\/
  contours are more flattened than the isophotes. There is a
  well-defined depression in the \hb\/ line strength within the central
  5\arcsec\/ \citep[see also][]{dav93} which we partly ascribe to
  imperfect removal of strong \hb\/ emission.
  
\item[\bf NGC\,4374 (M84):] This well known giant elliptical shows
  noticeable dust absorption \citep{qui00} and has a BL Lac nucleus
  \citep{bow00}. The metal lines have a similar morphology to those in
  NGC\,4278, following the isophotes in general. The \hb\/ absorption
  map is rather flat, showing a slight increase towards the edge of the
  field, also present in the data of \citet{dav93}, \citet{car93} and
  \citet{san04}.
  
\item[\bf NGC\,4382 (M85):] This well known object shows a distinct
  depression in \mgb\/ within the central 10\arcsec, consistent with
  \citet{fis96}, which appears slightly more flattened than the
  isophotes. This is coincident with an enhancement in the \hb\/
  absorption map \citep[see also][]{kun01}, as well as corresponding
  closely to a region of peculiar kinematic behavior (decoupled
  rotation, low velocity dispersion, and complex higher-order terms -
  see Paper~III). There is also evidence for a flattened feature in the
  iron absorption maps within this region, most evident in the Fe5015
  map.

\item[\bf NGC\,4387:] This boxy galaxy \citep{pel90} shows regular metal
  line gradients consistent with the isophotes. The \hb\/ map appears
  relatively flat in the inner regions while we observe a drop of \hb\/
  strength towards larger radii.
  
\item[\bf NGC\,4458:] This galaxy with a kinematically decoupled
  component (hereafter, KDC; see Paper~III) shows line strength
  gradients consistent with the isophotes. The region of the KDC is
  inconspicuous in the \hb\/ map \citep[see also][]{mor04}.
  
\item[\bf NGC\,4459:] This galaxy shows a central region of stronger
  \hb\/ absorption with a ring like structure around it. This is
  consistent with the central dust ring found in this galaxy; see
  Paper~V and \citet{sar01}. The metal lines show relatively regular
  gradients.
  
\item[\bf NGC\,4473:] This galaxy exhibits a complex morphology in the
  velocity dispersion map, with a region of high dispersion along the
  major-axis which widens at larger radii, which probably corresponds
  to two counter-rotating stellar disks. The metal line strength maps
  are more flattened than the isophotes, indicating a higher
  metallicity in the disk(s) while the \hb\/ map is rather flat.
  
\item[\bf NGC\,4477:] This galaxy shows a prominent misalignment of the
  kinematic and photometric major-axis. The metal line strength maps
  show an inner region with enhanced metallicity and show little
  structure elsewhere. There is weak evidence at larger radii along the
  major axis for increased metal line strength. The \hb\/ map appears
  flat.
  
\item[\bf NGC\,4486:] The line strength in the central 3\arcsec\/ and
  some part of the jet region of this famous Virgo galaxy are clearly
  affected by an imperfect removal of emission lines and non-thermal
  emission \citep[][]{dav93}. Overall, the line strength maps appear
  flat over large parts of the FoV. The \hb\/ map exhibits notably low
  values.
  
\item[\bf NGC\,4526:] This S0 object has a prominent dust disc which is
  not only visible in the reconstructed \sauron\ image but also clearly
  seen in the \mgb\/ and \hb\/ maps. The strong \hb\/ line strengths in
  the region of the dust lane indicates the presence of dust enshrouded
  young stars.
  
\item[\bf NGC\,4546:] This galaxy exhibits ionized gas counter-rotating
  with respect to the stars (see Paper~V). The \mgb\/ contours are more
  elongated in the direction of the rotation than the isophotes. This
  effect is less prominent in the iron maps. The \hb\/ map appears
  relatively flat with a hint of weaker \hb\/ absorption along the
  major axis.
  
\item[\bf NGC\,4550:] This S0 galaxy has two counter-rotating stellar
  disks of similar mass \citep{rix92}. There is evidence for the \mgb\/
  contours being flatter than the isophotes. The \hb\/ map shows a
  mildly elevated level in the central region along the major axis.
  
\item[\bf NGC\,4552:] This large elliptical shows regular line strength
  gradients, consistent with \citet{san04} and \citet{ram05}. The \hb\/
  map appears flat and shows notably low values.
  
\item[\bf NGC\,4564:] This highly elongated galaxy shows isoindex
  contours of the \mgb\/ and the Fe indices which are more flattened
  than the isophotes, indicating that the bulge and disk of this galaxy
  have different stellar populations.
  
\item[\bf NGC\,4570:] This edge-on S0 galaxy features multiple kinematic
  components such as an outer and a nuclear disk \citep{bosch98}. The
  \mgb\/ map and to a lesser extent the Fe maps show a central peak
  surrounded by a region depressed in line strength. At larger radii
  the onset of the outer disk, showing strong metal lines, can be seen.
  The \hb\/ map is relatively flat.
    
\item[\bf NGC\,4621:] This is another object with a strong indication
  from the stellar kinematics of a disc component, which is also seen
  in the metal line strength maps showing more flattened isoindex
  contours than the isophotes. The inner 60\,pc harbor a
  counter-rotating core \citep{wer02}. The \hb\/ map appears relatively
  flat.
  
\item[\bf NGC\,4660:] Similar to NGC\,4621, there are strong indications
  for a disk component in this galaxy, accompanied by pointy disky
  isophotes. This galaxy is among the best examples of \mgb\/ isoindex
  contours to be flatter than the isophotes for galaxies with strong
  rotation.
  
\item[\bf NGC\,5198:] This galaxy, with a central KDC, rotating nearly
  perpendicularly to the outer body, shows regular metal line gradients
  and a mild positive \hb\/ gradient.
  
\item[\bf NGC\,5308:] This is a typical case of a close to edge-on disc
  galaxy showing a rapidly rotating component (Paper~III) and a double
  sign reversal in $h_3$, indicating a central bar \citep{bur05}. The
  metal line strength gradients are regular and appear to be consistent
  with the isophotes.
  
\item[\bf NGC\,5813:] A galaxy with a well-known KDC
  \citep[][Paper~III]{Ef80,Ef82,ben94}. The line strength maps of this
  galaxy have already been presented in Paper~II. Here it is shown that
  we reproduce the large \mgb\/ gradient presented by \citet{gor90}.
  The KDC is not seen in the absorption line maps.
  
\item[\bf NGC\,5831:] The kinematics of this galaxy reveal the presence
  of a well-known KDC \citep[][Paper~III]{dav83, pel90}. The KDC is not
  really seen in the line strength maps, apart from the \mgb\/ map
  which appears to be more elongated in the direction of the KDC than
  the isophotes. The \hb\/ map appears relatively flat.
  
\item[\bf NGC\,5838:] This boxy galaxy is a strong rotator and
  \citet{pel90} show that the bulge of this galaxy contains old stellar
  populations with small color gradients, except for the inner
  2\arcsec. This is consistent with our line strength maps which show
  regular gradients and a mild peak of the \hb\/ map in the very
  center. There is evidence for the \mgb\/ contours being flatter than
  the isophotes. Our emission line maps (see Paper~V) also show
  significant detection in the center with evidence for a rotating
  structure.
  
\item[\bf NGC\,5845:] This galaxy is a compact elliptical, close to
  NGC\,5846. Small gradients in the line strength maps are seen,
  consistent with \citet{san04}. The central disk, clearly seen in the
  stellar kinematics (Paper~III), is not apparent in the line strength maps.
  
\item[\bf NGC\,5846:] The \sauron\/ field of view of this bright giant
  elliptical includes a foreground star and a companion (NGC\,5846A)
  north and south of its nucleus, respectively. The line strength maps
  are generally regular, with small gradients. The line strengths are
  consistent with \citet{san04} and \citet{ram05}.
 
\item[\bf NGC\,5982:] For this galaxy we confirm the presence of a KDC
  first detected by \citet{wag88}. The KDC is not clearly seen in the
  line strengths maps, however, there is evidence for the \mgb\/
  contours being flatter than the isophotes. The \hb\/ map appears
  relatively flat.
  
\item[\bf NGC\,7332:] Another strongly boxy galaxy with a KDC
  (counter-rotating) in the central 3\arcsec\ \citep[see][]{Fal04}.
  There is evidence for the \mgb\/ contours being flatter than the
  isophotes. The galaxy shows high \hb\/ values, indicating an age of
  about 3~Gyr \citep[see][]{Fal04}. The data are consistent with
  \citet{san04} and \citet{ram05}.
  
\item[\bf NGC\,7457:] This small S0 galaxy with a KDC shows surprisingly
  little structure in its line strength maps, consistent with the color
  profiles of \citet{pel99b}. The high \hb\/ values indicate a
  relatively young age.

\end{description}

\label{lastpage}
\end{document}